\documentclass[11pt,a4paper]{article}
\usepackage[english]{babel}
\usepackage{newlfont}
\usepackage{dsfont}
\usepackage{bm}
\usepackage[utf8x]{inputenc}
\usepackage{comment}
\usepackage{amsmath,amssymb}
\usepackage{mathtools,floatrow}
\usepackage{physics}

\usepackage[dvipsnames]{xcolor}
\usepackage{graphicx}

\usepackage{cite}
\numberwithin{equation}{section}
\numberwithin{table}{section}
\numberwithin{figure}{section}

\usepackage{caption}
\usepackage[normal]{subfigure}

\usepackage{hyperref}

\usepackage{tikz}
\usepackage{pgfplots}
\usepgfplotslibrary{groupplots}
\usetikzlibrary{decorations.markings, arrows}
\usepackage[margin=0.7in]{geometry}
\hypersetup{
	colorlinks = true,
	linkcolor = MidnightBlue,
	anchorcolor = blue,
	citecolor = OliveGreen,
	filecolor = blue,
	urlcolor = blue
}

\usepackage{float}
\usepackage{babel}
\usepackage{esint}
\usepackage{cite}

\definecolor{cbl}{rgb}{0,0,1}                

\newcommand{\bc}{\begin{center}}
\newcommand{\ec}{\end{center}}
\def\ba#1{\begin{array}{#1}\displaystyle}
\newcommand{\ea}{\end{array}}

\newcommand{\beq}{\begin{equation}}
\newcommand{\eeq}{\end{equation}}
\newcommand{\beqa}{\begin{eqnarray}}
\newcommand{\eeqa}{\end{eqnarray}}

\newcommand{\bi}{\begin{itemize}}
\newcommand{\ei}{\end{itemize}}

\newcommand{\ri}{{\mathrm i}}

\begin{document}

\begin{titlepage}
\vspace{0.2cm}
\begin{center}

{\large{\bf{The Staircase Model: Massless Flows and Hydrodynamics}}}

\vspace{0.8cm} 
{\large \text{Michele Mazzoni{ ${}^{\clubsuit}$}, Octavio Pomponio{${}^{\diamondsuit}$}, Olalla A. Castro-Alvaredo${}^{\heartsuit}$ and Francesco Ravanini${}^{\spadesuit }$}}

\vspace{0.8cm}
{\small
${}^{\clubsuit \diamondsuit\spadesuit}$ Dipartimento di Fisica, Universit\`a di Bologna, Via Irnerio 46, I-40126 Bologna, Italy\\
\vspace{0.2cm}
${}^{\heartsuit}$  Department of Mathematics, City, University of London, 10 Northampton Square EC1V 0HB, UK\\
\vspace{0.2cm}
${}^{\diamondsuit \spadesuit}$ INFN, Sezione di Bologna, Via Irnerio 46, I-40126 Bologna, Italy
}
\end{center}

\medskip
\medskip
\medskip
\medskip

\noindent The staircase model is a simple generalization of the sinh-Gordon model,  obtained by complexifying the coupling constant. This produces a new theory with many interesting features. 
Chief among them is the fact that scaling functions such as Zamolodchikov's $c$-function display ``roaming" behaviour, that is,  they visit the vicinity of an infinite number of  conformal fixed points, the unitary minimal models of conformal field theory. This rich structure also makes the model an interesting candidate for study using the generalized hydrodynamic approach to integrable quantum field theory (IQFT).  By studying hydrodynamic quantities such as the effective velocities of quasiparticles we can develop a more physical picture of interaction in the theory, both at and away from equilibrium. Indeed, we find that in the staircase model the effective velocity displays monotonicity features not found in any other IQFT.  These admit a natural interpretation when investigated in the context of the massless theories known as  $\mathcal{M}A_k^{(+)}$ models, which provide an effective description of massless flows between consecutive unitary minimal models. Remarkably, we find that the effective velocity in the staircase model can be reconstructed by a ``cut and paste" procedure involving the equivalent functions associated to the massless excitations of suitable $\mathcal{M}A_k^{(+)}$ models. We also investigate the average currents and densities of higher spin conserved quantities in the partitioning protocol.

\vfill

\noindent 
${}^{\clubsuit}$michele.mazzoni4@studio.unibo.it\\
${}^{\diamondsuit}$ octavio.pomponio@unibo.it\\
${}^{\heartsuit}$ o.castro-alvaredo@city.ac.uk\\
${}^{\spadesuit}$ francesco.ravanini@unibo.it\\

\hfill \today

\end{titlepage}

\section{Introduction}
Our understanding of the out-of-equilibrium dynamical properties of many-body quantum systems has vastly expanded over the past decade \cite{20,Eisert}. In the context of 1+1D quantum integrable models, a lot of interest was triggered by the results of the quantum Newton's cradle experiment \cite{kinoshita} which showed that dimensionality in conjunction with integrability give rise to a distinct kind of dynamics, one in which there is no long-term thermalization. This result was later related to the presence of infinitely many conserved quantities in integrable models. The dynamics is then determined by all conserved quantities, leading to the concept of generalized Gibbs ensembles (GGEs) \cite{Rigol}. As a consequence, quantum integrable models do not thermalize to Gibbs ensembles but they equilibrate to GGEs. 
Thanks to this fundamental understanding,  non-equilibrium dynamics has become a  powerful way of studying strongly correlated many-body systems \cite{Eisert,NERev2,NERev3,NERev4,NERev5,NERev6,NERev7,Brev,CEM}.

Generalized hydrodynamics (GHD) is one of the leading approaches to computing dynamical quantities in non-equilibrium steady states and non-stationary settings \cite{ourhydro,theirhydro,Brev}. The basic idea is that hydrodynamics emerges as a consequence of local entropy maximization and scale separation on individual fluid cells containing a mesoscopic quasi-particle number.  In a practical sense GHD allows us to evaluate exact expectation values of currents in GGEs, by employing the formula derived in \cite{ourhydro} within integrable quantum field theory (IQFT) and numerically checked in \cite{theirhydro} in quantum chains. This current formula has been subject to increasingly rigorous and general derivations \cite{cur1,cur2,cur3,dNBD2,cur4,cur5,cur6,cur7,cur8,cur9}.

The original GHD proposals considered  the partitioning protocol. In this set up two independently thermalized systems are put into contact at time zero. 
The presence of multiple conserved quantities gives rise to ballistic transport, meaning that, after a transient period, steady state currents flowing between the right and left sub-systems emerge; see the reviews \cite{NERev5,NERev3}. As mentioned above, GHD provides a method  to compute such currents by combining the hydrodynamic principle, generalized to infinitely many conservation laws, with an effective description of quasi-particles, which for IQFTs is based on the  thermodynamic Bethe ansatz (TBA) \cite{tba1,tba12,tba2}. The energy current and density in the partitioning protocol admit simple exact expressions in conformal field theory (CFT) \cite{EFlow,CFTcur2, CFTcur}. These constitute a useful benchmark when studying IQFTs, whose ultraviolet limits are described by CFT. 

The GHD framework has been generalized in many ways, to adapt to increasingly complex physical situations such as force terms \cite{DY,bastianello2019Integrability,BasGeneralised2019}, diffusive and higher corrections \cite{dNBD,GHKV2018,dNBD2,FagottiLocally}, noise \cite{bastianello2020noise} and integrability-breaking terms \cite{CaoTherm18,vas19,DurninTherma2020}. It has been shown to be superior to conventional hydrodynamics in describing the results of experiments on an atom chip \cite{chippy} and to qualitatively reproduce the phenomenology of Newton's cradle experiment \cite{cradle}. Impressive as these developments are, 
an aspect of the original formalism that has only become apparent recently is that, even when used in equilibrium situations, the hydrodynamic picture sheds new light into the physics of IQFTs. Put another way, well-known equilibrium features can be analysed from a different viewpoint by looking at hydrodynamic functions, such as the effective velocities of quasi-particles. This has become apparent in two recent studies of IQFTs with unstable particles in their spectrum \cite{ourunstable,tails}. These studies have shown that the presence of unstable particles in the theory, even if they are not part of the asymptotic particle spectrum, leads to recognizable signatures in the effective velocity profile and particle density of their stable constituents. One may for instance identify a particle density ascribable to the unstable particles. 
In this paper we apply GHD to the study of the staircase model (SM), introduced by Al. B. Zamolodchikov \cite{roaming} by generalizing the sinh-Gordon IQFT scattering matrix. The SM has features which make it comparable to the model studied in \cite{ourunstable,tails}. Namely, similar to the case of unstable particles, the two-body scattering matrix of the theory has poles in the unphysical sheet and the presence of these extra poles has an enormous effect on the physics of the theory. In particular, the monotonicity properties of typical hydrodynamic quantities such as the effective velocity are markedly different from those of any model studied so far and, as is typical of the model, are heavily influenced by the interplay between the value of a coupling $\theta_0$ and the value of the temperature (in a Gibbs ensemble) or temperatures (in the partitioning protocol set-up). Various generalizations of the SM construction were proposed in \cite{DOREY_1993,GenDR}.

\medskip

This paper is organized as follows:  In Section 2 we introduce the SM and the thermodynamic Bethe ansatz approach. In Section 3 we review the main principles of the GHD approach in the context of integrable quantum field  theory. In Section 4 we investigate several hydrodynamic quantities in the SM at equilibrium and identify unique features associated to this model: the spectral particle density and spectral particle current have generally many local maxima while the effective velocity displays intricate square-wave patterns \cite{video}. We provide an explanation of these properties based on the $\mathcal{M}A_k^{(+)}$ model, which provides an effective description of massless flows between consecutive unitary minimal models. In Section 5 we extend our investigation to the out-of-equilibrium situation where we consider the partitioning protocol. We focus on the expectation values of conserved densities and currents, both at and away from equilibrium. We evaluate these for several values of spin higher than 1 and observe that, under appropriate scaling and at sufficiently high temperatures, they too develop a staircase pattern. For the same higher spin currents/densities we investigate the temperature dependence in the CFT limit. We conclude in Section 6. In Appendix A we review some standard results for the SM following the derivations presented in  \cite{DOREY_1993,GenDR}. In particular, we present the constant TBA analysis that allows us to obtain equations for the height of individual plateaux of the TBA $L$-functions and derive the central charges of unitary minimal models. In Appendix B we present an explicit computation of higher spin currents in CFT as derived from a thermal TBA in the Ising model. We see that even for such a simple theory, the dependence on spin is relatively intricate.

\section{The Staircase Model}
\subsection{The Model}

The staircase model is constructed starting with the sinh-Gordon model. The sinh-Gordon model is the simplest interacting IQFT in one spatial dimension. It is defined by the  
 lagrangian
 \cite{toda2,toda1}:
\begin{equation}
   { \mathcal{L}}_{\mathrm{ shG}}=\frac{1}{2}(\partial_{\mu}\phi)^2-\frac{m^2}{g^2}\cosh(g\phi), \label{sg}
\end{equation}
where $\phi$ is the sinh-Gordon field, $m$ is the bare mass of the single particle in the spectrum and $g$ is the coupling constant.
 The two-particle scattering matrix associated to this model is given by
\cite{SSG2,SSG3,SSG} 
\begin{equation}\label{smatrix}
    S(\theta)=\frac{\tanh\frac{1}{2}\left(\theta-\frac{\ri\pi B}{2}\right)}{\tanh\frac{1}{2}\left(\theta+\frac{\ri\pi B}{2}\right)}.
\end{equation}
The parameter $B \in [0,2]$ is the effective coupling constant
which is related to the coupling constant $g$ in the
lagrangian by \cite{za}
\begin{equation}\label{BB}
    B(g)=\frac{2g^2}{8\pi + g^2}.
\end{equation}
The $S$-matrix is obviously
invariant under the transformation $B\mapsto 2-B$, a symmetry
which is also referred to as weak-strong coupling duality, as it
corresponds to $B(g)\rightarrow B(g^{-1})$ in (\ref{BB}).
The point $B=1$ is known as the self-dual point. 

In \cite{roaming} it was observed that when analytically continuing the coupling constant $B$ from the self-dual point to the complex plane via the transformation
\beq
B\mapsto 1+i\frac{2\theta_0}{\pi}\qquad {\mathrm{with}} \qquad \theta_0\, \in\, {\mathbb{R}}^+\,,
\label{trans}
\eeq
the resulting scattering matrix
\begin{equation}\label{sroaming}
    S(\theta)={\tanh\frac{1}{2}\left(\theta-\theta_0-\frac{i\pi}{2}\right)}{\tanh\frac{1}{2}\left(\theta+\theta_0-\frac{i\pi}{2}\right)}\,,
\end{equation}
still satisfies all physical requirements, such as unitarity and crossing, while being distinct from the sinh-Gordon $S$-matrix. In particular, it has poles in the unphysical sheet at $\theta=\pm\theta_0-\frac{i\pi}{2}$. In order to investigate the properties of this new theory, the paper \cite{roaming} carried out a detailed TBA study of the model and discovered
several unusual features. These observations were later generalized to the whole family of affine Toda field theories\cite{DOREY_1993,GenDR}, of which the sinh-Gordon model is the simplest example. We review the key ideas in the next subsection.

\subsection{Thermodynamic Bethe Ansatz and Scaling Function}
 \label{sec:staircase}
The TBA approach provides a description of IQFTs at finite temperature \cite{tba1,tba12,tba2}. This description can be easily generalized to GGEs, as discussed in \cite{Mossel} (even though the generalization had already been used in \cite{FM}).
The TBA equations describing the SM at thermal equilibrium take a very simple form. In fact, given that there is a single particle in the theory and that scattering is diagonal, there is a single equation to start with.
The non-trivial interaction that is present in the model enters the TBA equation through the kernel or two-body scattering phase $\varphi(\theta)$ which is defined as
\begin{equation}\label{eq:kernel}
	\varphi(\theta) = -i \frac{d}{d\theta}\log S(\theta) = \frac{1}{\cosh(\theta-\theta_0)} + \frac{1}{\cosh(\theta+\theta_0)}\,.
\end{equation}
This is a function that is peaked around $\theta=\pm \theta_0$. This means that interaction is maximal when the rapidity difference between interacting particles takes these values and greatly reduced otherwise.
 This feature plays a key role in the shape of some of the functions we will study later in the paper, specially the spectral particle density and spectral  particle current. 

For finite temperature $T = 1/\beta$ the pseudoenergies $\varepsilon(\theta)$ satisfy the nonlinear integral equation (TBA-equation) 
\begin{equation}\label{eq:TBAstair}
	\varepsilon(\theta) = \beta \cosh\theta - (\varphi \star L)(\theta),
\end{equation}
where we have taken the mass $m=1$ and $\star$ denotes the rapidity convolution
\begin{equation}
	(\varphi \star L)(\theta) = \frac{1}{2\pi}\int d\theta' \varphi(\theta-\theta')L(\theta').
\end{equation}
Since $S(0)=-1$ the TBA system is of fermionic type \cite{tba1,tba12,tba2} and therefore
\begin{equation}
	L = \ln(1+e^{-\varepsilon(\theta)}).
\end{equation}
The ground state energy is given by 
\begin{equation}
	E(\beta) = -\frac{1}{2\pi}\int L(\theta)\cosh\theta  \,d\theta =  -\frac{\pi c(y)}{6\beta}\,,
	\label{Ecy}
\end{equation}
where $c(y)$ is the TBA scaling function and
\beq
 y = \ln\frac{2}{\beta}
 \label{they}
 \eeq
 is a new, more convenient variable in terms of which the infrared (IR) or low temperature limit corresponds to $y\rightarrow -\infty$ and the ultraviolet (UV) or high temperature limit corresponds to $y\rightarrow \infty$.
 The scaling function flows from the UV value $c(\infty)=1$ corresponding to the central charge of a free massless boson, to the IR value $c(-\infty)=0$. So far, this behaviour is identical to that of the sinh-Gordon theory.
The unusual features of the model become only apparent when the dependence of the $L$-function and the scaling function on the parameter $\theta_0$ is explored in detail. These have been discussed in the original paper \cite{roaming} and in \cite{DOREY_1993,GenDR} but it is worth recalling the main ideas, as they will play a role in the hydrodynamic picture.

Let us examine the change in the function $L(\theta)$ as we increase $y$ towards the deep UV regime.  We take $\theta_0 \gg 1$ in order to have two well separated bumps in the kernel (\ref{eq:kernel}).  We have the following regimes:

\begin{itemize}

\item  For $y>0$, $L(\theta)$ develops a plateau of height $\log2$ in the region $-y<\theta<y$, that is the usual free fermion UV behaviour. As $y$ increases the plateau broadens. 
At the same time, the scaling functions $c(y)$ reaches the value  $c=1/2$ corresponding to a massless free fermion or the Ising model. This is the first of the series of unitary minimal models and we denote it by $\mathcal{M}_3$ for reasons that become clear below. 
This $L$-function is shown in the top-right of Fig.~\ref{fig:collective_1}.
\item When $y$ reaches the value $\theta_0/2$, the plateau's edges start interacting via the kernel in equation (\ref{eq:TBAstair}). The interaction can be described by two separate interacting integral equations for two separate pseudoenergies $\varepsilon_{\pm}(\theta)=\varepsilon(\theta\pm \frac{\theta_0}{2})$. The resulting system coincides with the description of the massless flow between the critical ($\mathcal{M}_3$) and tricritical  ($\mathcal{M}_4$) Ising models given in \cite{ZAMOLODCHIKOV1991524}. This relation is explored in later sections. 
As $y$ increases the scaling function grows from $c=1/2$ to $c=7/10$, as expected, while the $L$-function develops two higher plateaux at values $2\log(2\cos\frac{\pi}{5})$, which rise above the main plateau at $\log 2$. This $L$-function is shown in the top-right of Fig.~\ref{fig:collective_2}. This specific flow has also being studied using form factor techniques in \cite{DDG}.

\item Similarly, when $y$ reaches $\theta_0$ we have a new transition from $c=7/10$ to $c=4/5$ which again describes the massless flow between two unitary minimal models, in this case the  tricritical and the tetracritical Ising model ($\mathcal{M}_5$). This $L$-function is shown in the top-right of Fig.~\ref{fig:collective_3}.

\begin{figure}[h!]
\floatbox[{\capbeside\thisfloatsetup{capbesideposition={right,top},capbesidewidth=4cm}}]{figure}[\FBwidth]
{\caption{Scaling function of the SM for $\theta_0=20,35,50$. The central charges $c_{k+2}$ (dashed horizontal lines for $k=1,2,3,4,5$) are approached by $c(y)$ at different values of $y$, as the crossover points are at $y \sim k\theta_0/2$.}\label{fig:scaling_function}}
{
	\includegraphics{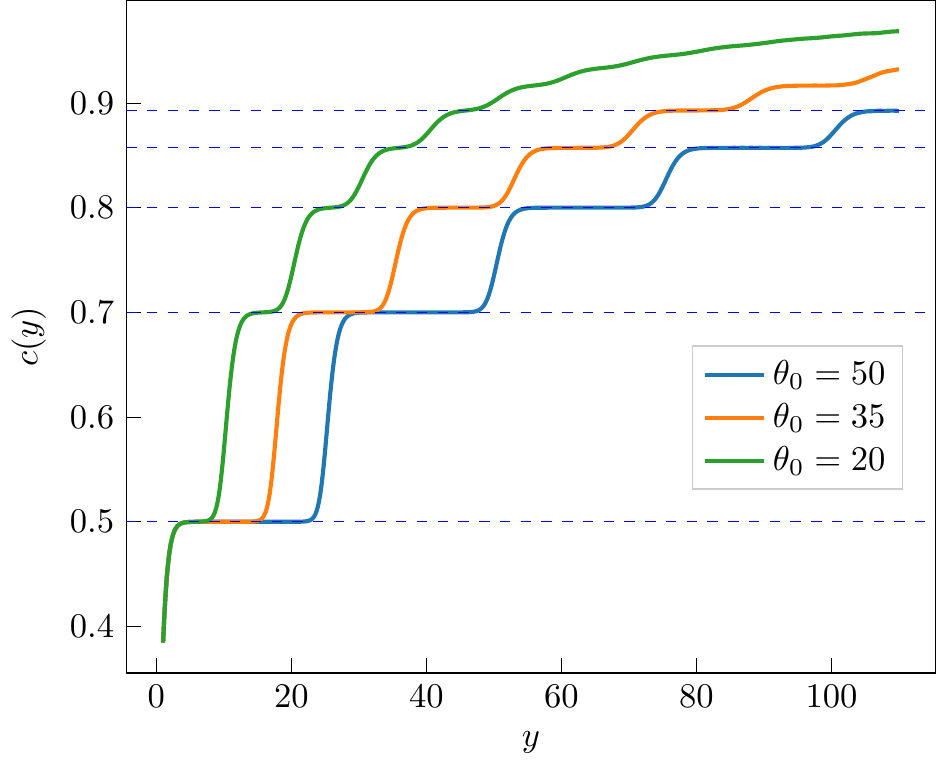}}
	\end{figure}
	
\item  In general, every time $y\sim k \theta_0/2$, with $k=1,2,3,\dots$ the scaling  function flows (or roams) from the value
\begin{equation}
	c_{k+2} = 1-\frac{6}{(k+2)(k+3)}, \qquad k=1,2,\ldots
	\label{cp}
\end{equation}
that is, the central charge of $\mathcal{M}_{k+2}$, to the value $c_{k+3}$ corresponding to $ \mathcal{M}_{k+3}$. Simultaneously, the $L$-function acquires new plateaux, emerging symmetrically in $\theta$ in pairs.
In summary, for $\theta_0 \gg 1$ the function $c(y)$ exhibits  a staircase behaviour roaming through the value $c_{k+2}$ inside every interval $(k-1)\theta_0/2<y<k\theta_0/2$ and finally settling on the UV value $c=\lim_{k\rightarrow \infty} c_k=1$. The scaling function $c(y)$ is depicted in Fig.~\ref{fig:scaling_function} for three different values of the parameter $\theta_0$.
\end{itemize}

The values (\ref{cp}) are not only observed numerically when evaluating $c(y)$ through the definition (\ref{Ecy}), but they can also be obtained analytically by studying the constant TBA system associated with the SM. That is, the set of TBA equations that are obtained in the UV limit. This is discussed in detail in Appendix A.

\subsection{ $\mathcal{M}A_k^{(+)}$ Massless Flows}
\label{Anseries}
As we have seen, the SM is associated with an RG flow which visits the vicinity of all unitary minimal models. 
In particular, each pair of successive fixed points $\mathcal{M}_{k+2}$ and $\mathcal{M}_{k+1}$ are connected by RG trajectories corresponding to a massless interpolating field theory whose UV limit is $\mathcal{M}_{k+2}$, while the IR limit is controlled by $\mathcal{M}_{k+1}$. The associated RG flow is generated from the UV fixed point by  perturbing with the relevant operator $\Phi_{13}$ and negative coupling \cite{ZAMOLODCHIKOV1991524, ZAMOLODCHIKOV1991497,onemore}. More details can be found at this end of this Section. 

\begin{figure}[h!]
	\centering
	\begin{tikzpicture}
	
		\draw[thick] (0,0) [] circle(0.15);
			\node[] at (0.05, 0.5) {$\frac{\beta}{2}e^{-{\theta}}$};
			\node[] at (0, -0.5) {$1$};
		\draw[thick] (.15,0) -- (1.35,0);
		\draw[thick] (1.5,0) [] circle(.15);
					\node[] at (1.5, -0.5) {$2$};
		\draw[thick] (1.65,0) -- (1.9,0);
		\draw[line width=0.28mm, densely dotted] (1.9,0) -- (2.15,0);
		\draw[line width=0.28mm, densely dotted] (3,0) -- (3.2,0);
		\draw[thick] (3.2,0) -- (3.45,0);
		\draw[thick] (3.6,0) [] circle(.15);
			\node[] at (3.6, -0.5) {$k-1$};
		\draw[thick] (3.75,0) -- (4.85,0);
		\draw[thick] (5,0) [] circle(.15);
			\node[] at (5, 0.5) {$\frac{\beta}{2}e^{{\theta}}$};
			\node[] at (5, -0.5) {$k$};
		
	\end{tikzpicture}
	\caption{$A_k$ Dynkin diagram showing driving terms associated to the massless TBA  (\ref{eq:TBAdyn}). }\label{fig:A_nDynkin}
\end{figure}
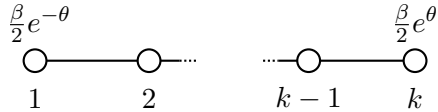

Much as the example discussed earlier of the flow between critical and tricritical Ising, which can be described by a pair of TBA equations for shifted pseudoenergies,
the interpolating flow between generic fixed points $\mathcal{M}_{k+2}$ and $\mathcal{M}_{k+1}$  can be described  by a set of TBA equations associated with the $A_{k}$ Dynkin diagram \cite{ZAMOLODCHIKOV1991524, ZAMOLODCHIKOV1991497,onemore}. They take the form
\begin{equation}
\label{eq:TBAdyn}
	\varepsilon_i(\theta) = \omega_i(\theta) - \sum_{j=1}^{k}(\hat{\varphi}_{ij}\star L_j)(\theta)\qquad {\mathrm{with}} \qquad \hat{\varphi}_{ij}(\theta) = \frac{I_{ij}}{\cosh\theta}\,
\end{equation}
and $I_{ij}=\delta_{i, j+1}+ \delta_{i,j-1}$ is  the adjacency matrix of the $A_k$ algebra (see Fig.~\ref{fig:A_nDynkin}). The driving terms $\omega_i(\theta)$ are given by
\begin{equation}
	\omega_i(\theta) = \frac{\beta}{2}(e^{-\theta}\delta_{i,1} + e^{\theta}\delta_{i,k}).
\end{equation}
Note that the driving terms associated to nodes $1$ and $k$ are usually exchanged in the literature, but this is a more suitable prescription in order to relate the $\mathcal{M}A_k^{(+)}$ model to the SM.

In the $\mathcal{M}A_k^{(+)}$ model, each node in the Dynkin diagram and each TBA equation can be interpreted as associated to a different particle species,  with non-trivial scattering only between nearest neighbouring nodes. Only two nodes in the Dynkin diagrams couple to non-vanishing driving terms. The source
terms $\frac{\beta}{2}e^{\pm\theta}$ correspond to right- and left-moving i.e. (RM) and (LM) particles, whose one-particle energy and momentum are 
\beq
e_{\pm}(\theta)=e^{\pm \theta-y}\quad {\mathrm{and}} \quad p_{\pm}(\theta)=\pm e^{\pm \theta-y}\,,
\eeq
  respectively ($y$ is the variable defined in \eqref{they}). The other nodes in the Dynkin diagram correspond to magnons, which describe internal degrees of freedom of the quasi-particle excitations. It is important to note that although the magnonic excitations themselves may be regarded as quasi-particles, they have zero one-particle eigenvalues with respect to the conserved charge operators, hence quantities such as the energy or momentum are carried only by the RM and LM species. In particular the scaling function
\begin{equation}
	c(y) = \frac{3 \beta}{\pi^2} \sum_{j=1}^{k} \int_{-\infty}^{\infty} d \theta\, p'_j(\theta) L_j(\theta)\,,
	\label{scaling}
\end{equation}
only receives equal contributions from the LM and RM. A GHD study of this and other massless theories was carried out in \cite{DXHAk}.

It is worth saying a little more about the origins of this massless model.  The $\mathcal{M}A_k^{(+)}$ models  can be seen as  perturbations of $\mathcal{M}_{k+2}$ by the field $\Phi_{13}$ ($k=1,2,3 \ldots$). 
There are however two families of theories with different properties depending on whether or not the perturbing parameter is positive. These two families were named  $\mathcal{M} A_k^{(\pm)}$ in \cite{ZAMOLODCHIKOV1991497}. 

The family $\mathcal{M} A_k^{(-)}$ is massive, and the $S$-matrix
 has been written by various authors \cite{restrict1, restrict2, ZAMOLODCHIKOV1991497}. It corresponds to a restriction of the sine-Gordon model (that is, the sine-Gordon model at a specific $k$-dependent value of the coupling constant) which is also related to the restricted solid-on-solid (RSOS)$_k$ lattice models. The scattering in these theories is non-diagonal but the associated TBA system can be diagonalized resulting into an $A_k$-based system of $k$ TBA equations where the driving terms are $\omega_1(\theta)=\cosh\theta$, $\omega_i(\theta)=0$ for $i>1$. Thus the $A_k$-system in this case describes a massive particle and $k-1$ magnons. 
 
The perturbation of interest in the context of the SM is $\mathcal{M} A_k^{(+)}$ which was studied in \cite{ZAMOLODCHIKOV1991524}. These models are massless (although non conformal, there is a fundamental scale in the theory) and describe a crossover between the unitary minimal models $\mathcal{M}_{k+2}$ and $\mathcal{M}_{k+1}$. This flow was studied perturbatively for large $k$ some years before \cite{sovzam,CarLud}. The TBA description which was then conjectured is given by the equations (\ref{eq:TBAdyn}). 

\subsection{SM $L$-function from the $\mathcal{M}A_k^{(+)}$ Model}
\begin{figure}[h!]
	\centering
	{
		\includegraphics{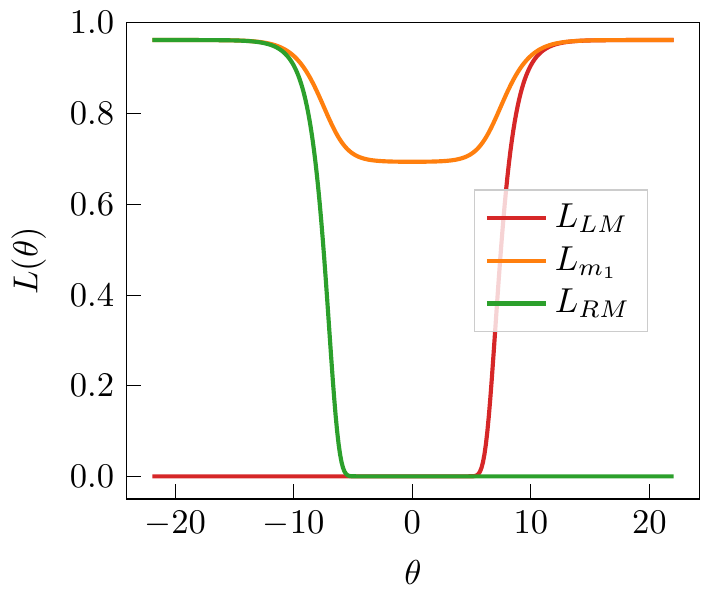} 
		\includegraphics{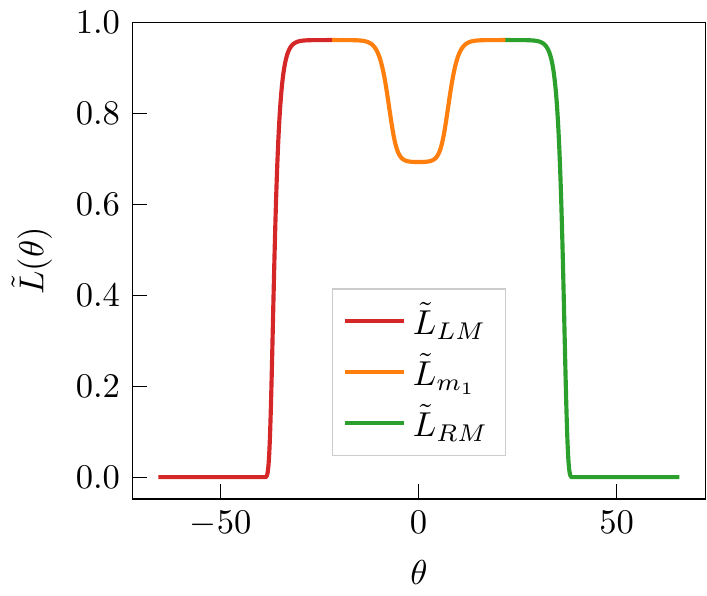}
	}
	\caption{\emph{Left}: $L$-functions for the $\mathcal{M}A_3^{(+)}$ model, showing solutions for three functions corresponding to the RM, the LM and the magnon. \emph{Right}: Same functions shifted in rapidity space to reconstruct the single $L$-function of the SM.} \label{fig:comp}
\end{figure}

A remarkable feature of the effective description of massless flows provided by the $\mathcal{M}A_k^{(+)}$ model is that
it is possible to carry out both a quantitative and  qualitative comparison between solutions of (\ref{eq:TBAstair}) and (\ref{eq:TBAdyn}) by introducing ad hoc shifts (in $\theta$ space) of the solutions of (\ref{eq:TBAdyn}). This procedure amounts to ``cutting an pasting" the solutions for the $L$-functions of the $\mathcal{M}A_k^{(+)}$ model in such a way as to reproduce the SM $L$-function  corresponding to a fixed finite value of  $\alpha=\frac{y}{\theta_0}$ for $y,\theta_0 \rightarrow \infty$. An illustration of this approach is presented in Fig.~\ref{fig:comp}. As can be seen in the example, such engineered $L$-function clearly matches the $L$-function structure of the SM for $\frac{1}{2}< \alpha < 1$.

\begin{figure}[!ht]
	\centering		\begin{tikzpicture}
	\definecolor{color0}{rgb}{0.83921568627451,0.152941176470588,0.156862745098039}
\definecolor{color1}{rgb}{1,0.498039215686275,0.0549019607843137}
\definecolor{color3}{rgb}{0.172549019607843,0.627450980392157,0.172549019607843}
		\node at (-0.03,0.25) {\includegraphics{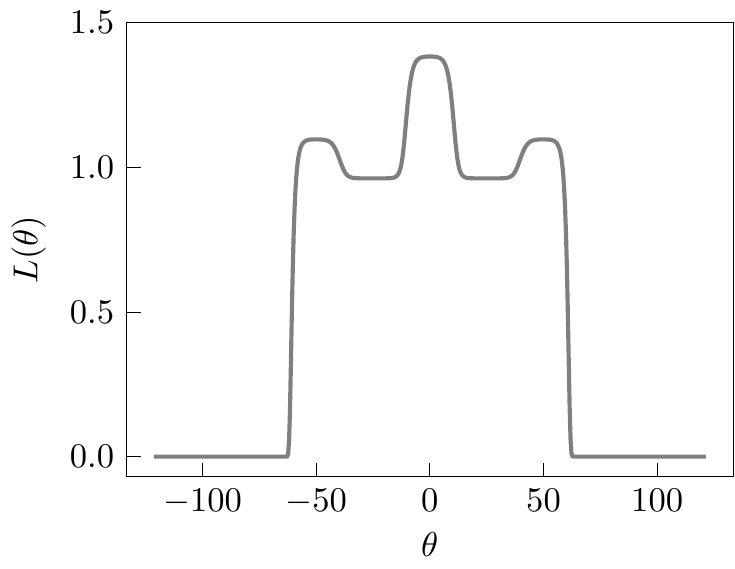}};
		
		\draw[thick, densely dotted] (-0.55,2.1) -- (-0.55,1.71);		
		\draw[thick, densely dotted] (0.62,2.1) -- (0.62,2.51);		
		\draw[thick, densely dotted] (1.75,2.1) -- (1.75,1.71);
		\draw[thick] (-0.6,2.1) -- (1.7,2.1);
		
		\draw[] (-0.55,2.1) [fill=color0]circle(0.15);
		\draw[] (0.62,2.1) [fill=color1]circle(0.15);
		\draw[] (1.75,2.1) [fill=color3]circle(0.15);

		\draw[thick, densely dotted] (-1.5,0) -- (-1.5,-1.46);		
		\draw[thick, densely dotted] (2.6,0) -- (2.6,-1.46);	
		\draw[thick, densely dotted] (-0.05,0) -- (-0.05,1.3);			
		\draw[thick, densely dotted] (1.25,0) -- (1.25,1.3);			
		
		\draw[thick] (-1.5,0) -- (2.6,0);
		\draw[] (-1.5,0) [fill=color0]circle(0.15);
		\draw[] (-0.05,0) [fill=color1]circle(0.15);
		\draw[] (1.25,0) [fill=MidnightBlue]circle(0.15);
		\draw[] (2.6,0) [fill=color3]circle(0.15);
		
	\end{tikzpicture}
	
	\begin{subfigure}
		{\includegraphics{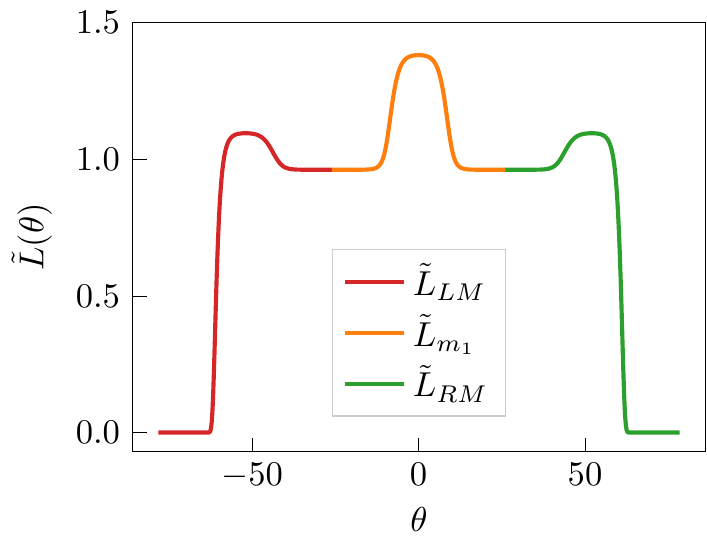}}
	\end{subfigure}
	\begin{subfigure}
		{\includegraphics{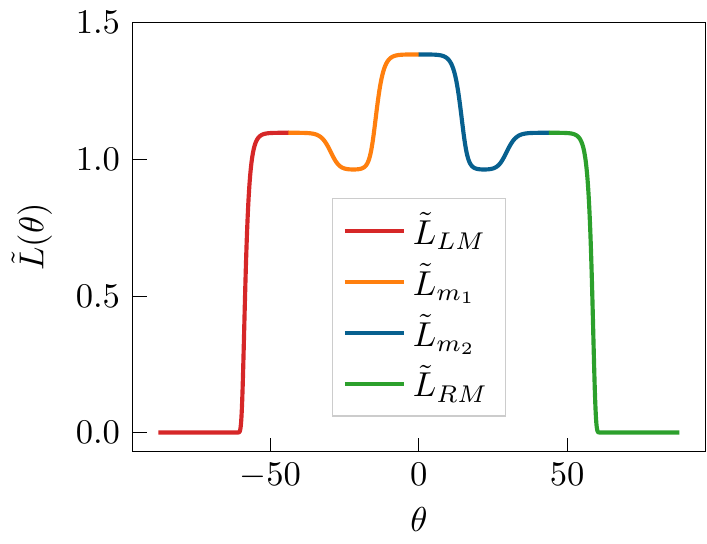}}
	\end{subfigure}
	
\caption{\emph{Top}: $L$-function of the SM for $1<\alpha<3/2$ ($k=3$) highlighting the $2k+1=7$ plateaux' mid-points $z_i$  and $2k$ kinks $K_i$ as discussed in Appendix A.  \emph{Bottom, Left}: $L$-functions of the $\mathcal{M}A_3^{(+)}$ model in the UV limit, reproducing the 
odd-labelled plateaux of the SM model and matching the colour scheme of the top panel. \emph{Bottom, Right}: $L$-functions of the $\mathcal{M}A_4^{(+)}$ model in the IR limit, reproducing the even-labelled plateaux of the SM model and matching the colour scheme of the top panel. }\label{fig:limits}
\end{figure}

This procedure allows us to see each step of the scaling function of the SM as two different limits, encoded in the two sets of decoupled equations (\ref{eq:sets}) for the values of the pseudoenergy at the centres of the plateaux found in Appendix A: if the RG trajectory flows very close to the fixed point $\mathcal{M}_{k+2}$, the latter can be seen both as the UV limit of the $A_{k}$ flow and as the IR limit of the $A_{k+1}$ flow of the massless TBA (\ref{eq:TBAdyn}). This is illustrated in Fig.~\ref{fig:limits}. 

\section{Generalized Hydrodynamics: A Short Review}
\label{GHDintro}
The main equations of GHD have been reviewed in many places (see i.e. \cite{Brev}). Here we will focus just on the key equations and on the functions that we will be examining in the paper.
GHD is based upon a description of quasi-particles which is provided by the TBA formulation. We have already written the TBA equation for the SM in (\ref{eq:TBAstair}). However, a more general version of this equation takes
the form
\begin{equation}\label{eq:TBAgge}
	\varepsilon(\theta) = \sum_i \beta_i h_i(\theta) - (\varphi \star L)(\theta),
\end{equation}
where $h_i(\theta)$ are the one-particle eigenvalues of the conserved quantities in the model. For instance $h_0(\theta)=1$ (particle number), $h_1(\theta)=e(\theta)=\cosh\theta$ (energy), $h_2(\theta)=p(\theta)=\sinh\theta$ (momentum) and more generally
\beq
h_{2s-1}(\theta)=\cosh(s\theta) \qquad h_{2s}=\sinh(s\theta), \qquad {\mathrm{for}} \quad s\in \mathbb{Z}_{\geq 0}\,.
\label{onep}
\eeq
Recall that we have set the mass scale $m=1$. 
The parameters $\beta_i$ are generalized inverse temperatures. A new equation is obtained by differentiating (\ref{eq:TBAgge}) with respect to one of these parameters, say $\beta_i$. The resulting ``dressing" equation 
can be written as
\beq
h_i^{\mathrm{dr}}(\theta)=h_i(\theta)+(\varphi \star g_i)(\theta), \qquad \mathrm{with} \qquad g_i(\theta)=n(\theta) h_i^{\mathrm{dr}}(\theta)
\label{dressing}
\eeq
and is a linear equation for the dressed quantities $h^{\mathrm{dr}}(\theta)$. The occupation functions $n_i(\theta)$ are defined as
\begin{equation}\label{eq:occ}
	n_i(\theta) = \frac{1}{1+e^{\varepsilon_i(\theta)}}=1-e^{-L_i(\theta)}.
\end{equation}

These dressed quantities enter the formulae 
\beqa 
\texttt{q}_i =  \int_{-\infty}^\infty \frac{d p(\theta)}{2\pi} h_i^{\mathrm{dr}}(\theta) n(\theta)=
 \int_{-\infty}^\infty \frac{d \theta}{2\pi} e(\theta) h_i^{\mathrm{dr}}(\theta) n(\theta)= 
 \int_{-\infty}^\infty \frac{d \theta}{2\pi} e^{\mathrm{dr}}(\theta) h_i(\theta) n(\theta)\,,
\label{q}
\eeqa 
and 
\beqa 
\texttt{j}_i =  \int_{-\infty}^\infty \frac{d e(\theta)}{2\pi} h_i^{\mathrm{dr}}(\theta) n(\theta)=
\int_{-\infty}^\infty \frac{d \theta}{2\pi} p(\theta) h_i^{\mathrm{dr}}(\theta) n(\theta)
= \int_{-\infty}^\infty \frac{d \theta}{2\pi} p^{\mathrm{dr}}(\theta) h_i(\theta) n(\theta)\,,
\label{j}
\eeqa 
for the average density $\texttt{q}_i$ and the average current $\texttt{j}_i $ associated to the conserved quantity of one-particle eigenvalue $h_i(\theta)$. One of the main results of \cite{ourhydro,theirhydro} 
was the expression for the average current, which turned out to be related to the average density via crossing (the rigorous proof of this formula has been examined in many papers since  \cite{cur1,cur2,cur3,dNBD2,cur4,cur5,cur6,cur7,cur8,cur9}). The last equality in the equations above (i.e. the fact that the dressing can be applied to either $h_i(\theta)$ or to the energy/momentum without changing the result)
follows from the TBA- and dressing equations. 

At thermal equilibrium these equations are sufficient to describe all quantities of interest, some of which have a natural hydrodynamic interpretation. Besides the densities and currents, we are interested in the functions
\beqa \label{eq:charge}
  \rho_p(\theta)&=& e^{{\mathrm{dr}}}(\theta) n(\theta) \qquad \mathrm{spectral\, particle\, density},\\
   v^{{\mathrm{eff}}}(\theta)&=&\frac{(e')^{{\mathrm{dr}}}(\theta)}{(p')^{{\mathrm{dr}}}(\theta)} \qquad\,\,\, \mathrm{effective\,\, velocity}, \label{eq:velocity}\\
     \rho_c(\theta)&=&  \rho_p(\theta)v^{{\mathrm{eff}}}(\theta) \quad \,\, \mathrm{spectral\,particle\, current}\,.
\eeqa
The spectral particle density and spectral particle current are functions whose integration in $\theta$ gives the particle density $\texttt{q}_0$ and the particle current $\texttt{j}_0$. The effective velocity is the velocity of propagation of quasi-particles due to the presence of interaction. From the definition, $v^{{\mathrm{eff}}}(\theta)=\tanh\theta$ for free fermions/bosons (when the kernel is vanishing) but changes when interaction is present. In most IQFTs such as the sinh-Gordon model studied in \cite{ourhydro} this change is rather subtle. However, as we will see later, for the SM the change is enormous and is fully explained by the massless picture provided by the $\mathcal{M}A_k^{(+)}$ model. 

Finally, a word is due regarding the partitioning protocol. So far we have defined some hydrodynamic quantities for a GGE described by (\ref{eq:TBAgge}). However, the same equations (\ref{q}), ({\ref{j}}) and (\ref{eq:charge}) hold if
we instead want to describe the non-equilibrium steady state (NESS) that will be formed if two copies of the SM model are first thermalized at different inverse temperatures $\beta_L$ and $\beta_R$ and are then  connected at time $t=0$ and let to evolve for a long time. Typically a NESS will emerge in the central region between the two subsystems. Within this NESS there will be non-equilibrium ballistic currents emerging described by (\ref{j}). In \cite{ourhydro,theirhydro} a very simple characterization of this NESS was found. The NESS is fully determined by the occupation function $n(\theta)$ which takes the form
\beq
n(\theta)=n_R(\theta) \Theta(\xi- v^{{\mathrm{eff}}}(\theta;\xi))+n_L(\theta)\Theta( v^{{\mathrm{eff}}}(\theta;\xi)-\xi)\,,
\label{npar}
\eeq
where $n_{L/R}(\theta)$ are the occupation functions in the right/left reservoir and $ v^{{\mathrm{eff}}}(\theta;\xi)$ is the effective velocity, now also a function of the ray $\xi:=\frac{x}{t}$. 
In consequence, all functions defined above will also depend on the ray $\xi$.
They depend only on the ray and not on $x$ and $t$ independently because the problem is scale invariant. 

\begin{figure}[h!]
\centering
		{
	    \includegraphics{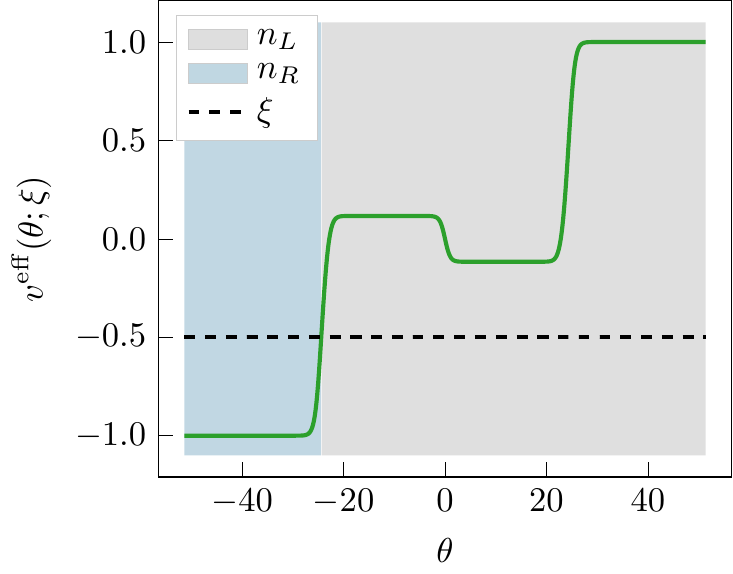}
		\includegraphics{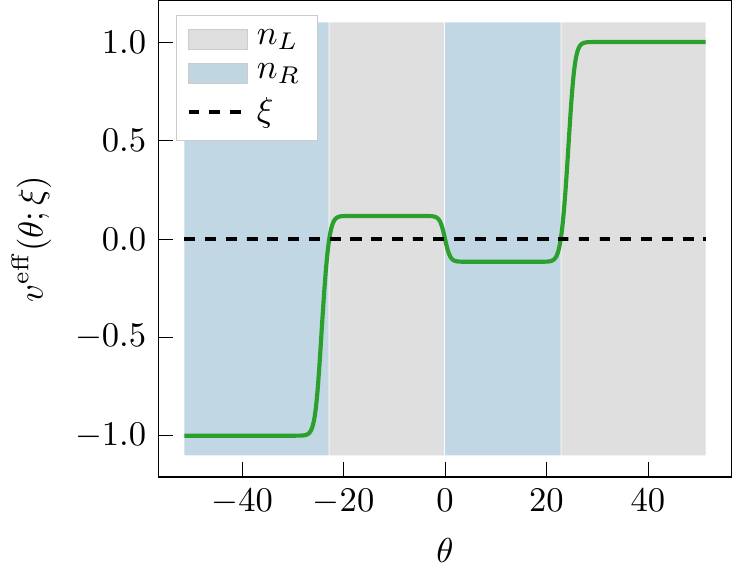}
		}
		\caption{A sketch of the effective velocity in the region $\frac{1}{2}<\alpha<1$ ($k=2$, second plateau of $c(y)$). As we can see, depending on the ray we can have multiple solutions to $v^{{\mathrm{eff}}}(\theta;\xi)=\xi$ and every time the velocity changes sign, so does $n(\theta)$ alternate between the right (blue) and left (gray) solutions.}
		\label{vefff}
	\end{figure}	
The structure above admits a simple interpretation:  particles on the right (left) of a given ray have come from the left (right) where would have been originally thermalized at inverse temperature $\beta_{L/R}$. 
A special feature of the SM is that the equation $ v^{{\mathrm{eff}}}(\theta;\xi)=\xi$ may have multiple solutions depending on $\xi$ and $\beta_{L/R}$, therefore the function $n(\theta)$ has multiple discontinuities as shown in Fig.~\ref{vefff}.

\section{The SM at Equilibrium}
\begin{figure}[ht!]
	\centering
	{
	\includegraphics{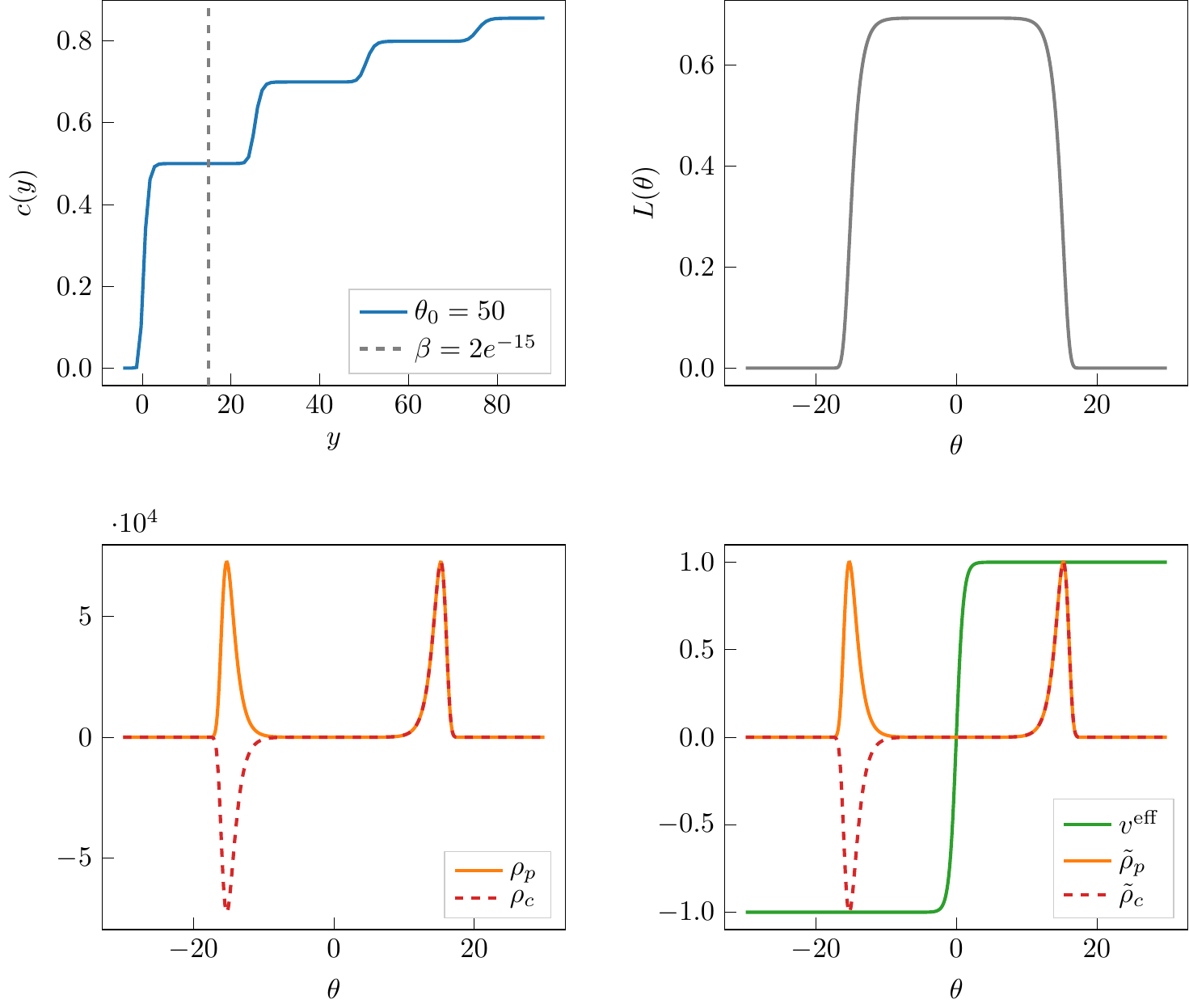}
	}
	\caption{Free fermion regime for $k=1$. \emph{Top}: Scaling function and $L$-function for $\beta = 2e^{-15}$ and $\theta_0=50$, hence $\alpha = 3/10<\frac{1}{2}$.
	 \emph{Bottom Left}: Spectral density (solid orange line) and  spectral particle current (dashed red line). \emph{Bottom Right}: Effective velocity (green) versus scaled spectral particle functions $\tilde{\rho}_{p/c}(\theta)$.
	 Since we are in the free fermion regime, the effective velocity is simply $\tanh\theta$ and the spectral particle densities exhibit two separate bumps corresponding to equal densities of right- and left-moving fermions.}
	 \label{fig:collective_1}
\end{figure}

\begin{figure}
	\centering
		{
	\includegraphics{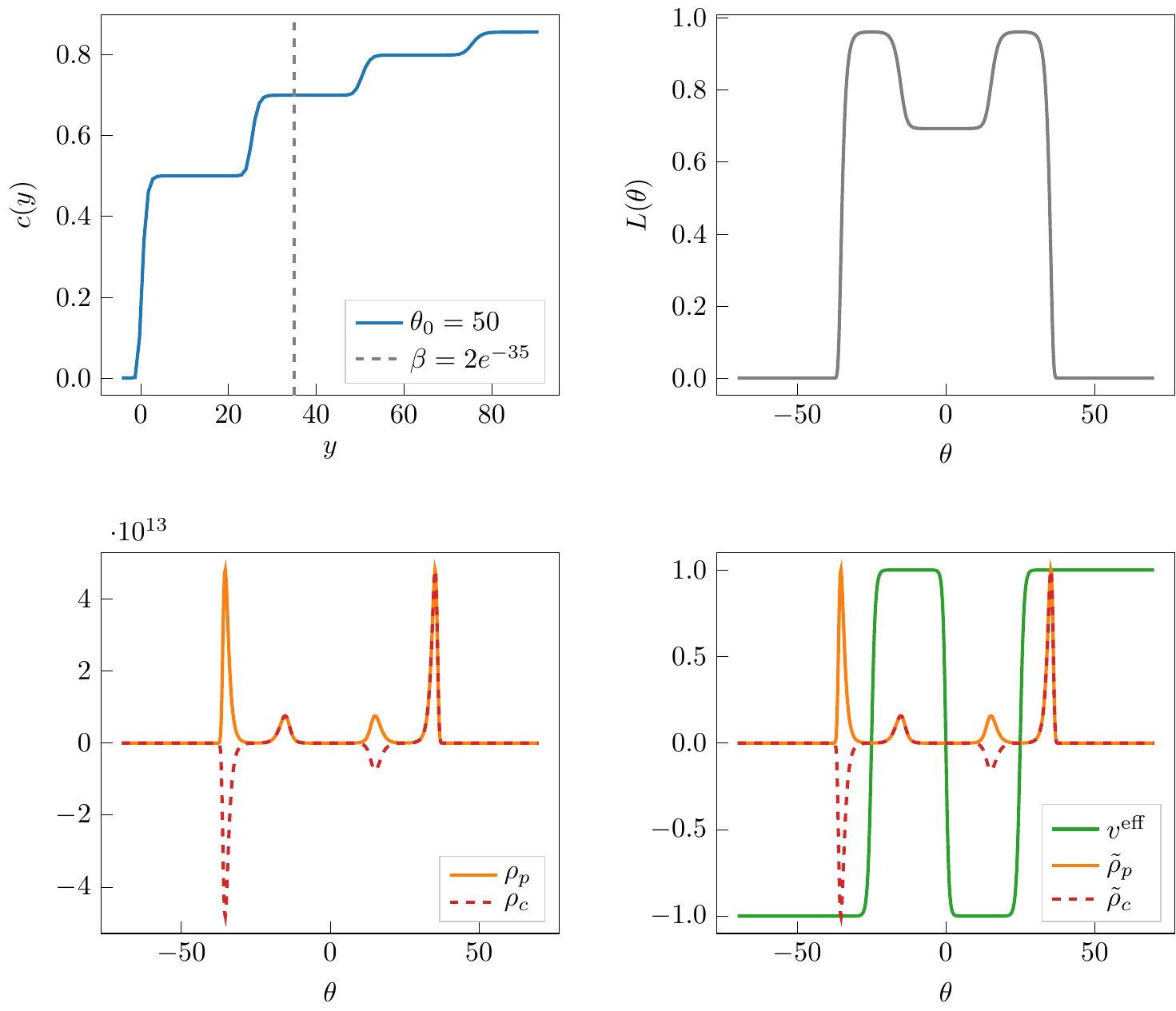}
	}
	\caption{Tricritical Ising for $k=2$: \emph{Top}: Scaling function and $L$-function for $\beta = 2e^{-35}$ and $\theta_0=50$, hence $\frac{1}{2}<\alpha = 7/10<1.$  \emph{Bottom Left}: Spectral density (solid orange line) and  spectral particle current (dashed red line). \emph{Bottom Right}: Effective velocity (green) versus scaled spectral particle functions (\ref{sca}). Note that the effective velocity is now non-monotonic and has $2k-1=3$ zeroes.}\label{fig:collective_2}
\end{figure}

\begin{figure}[h!]
	\centering
	{
	\includegraphics{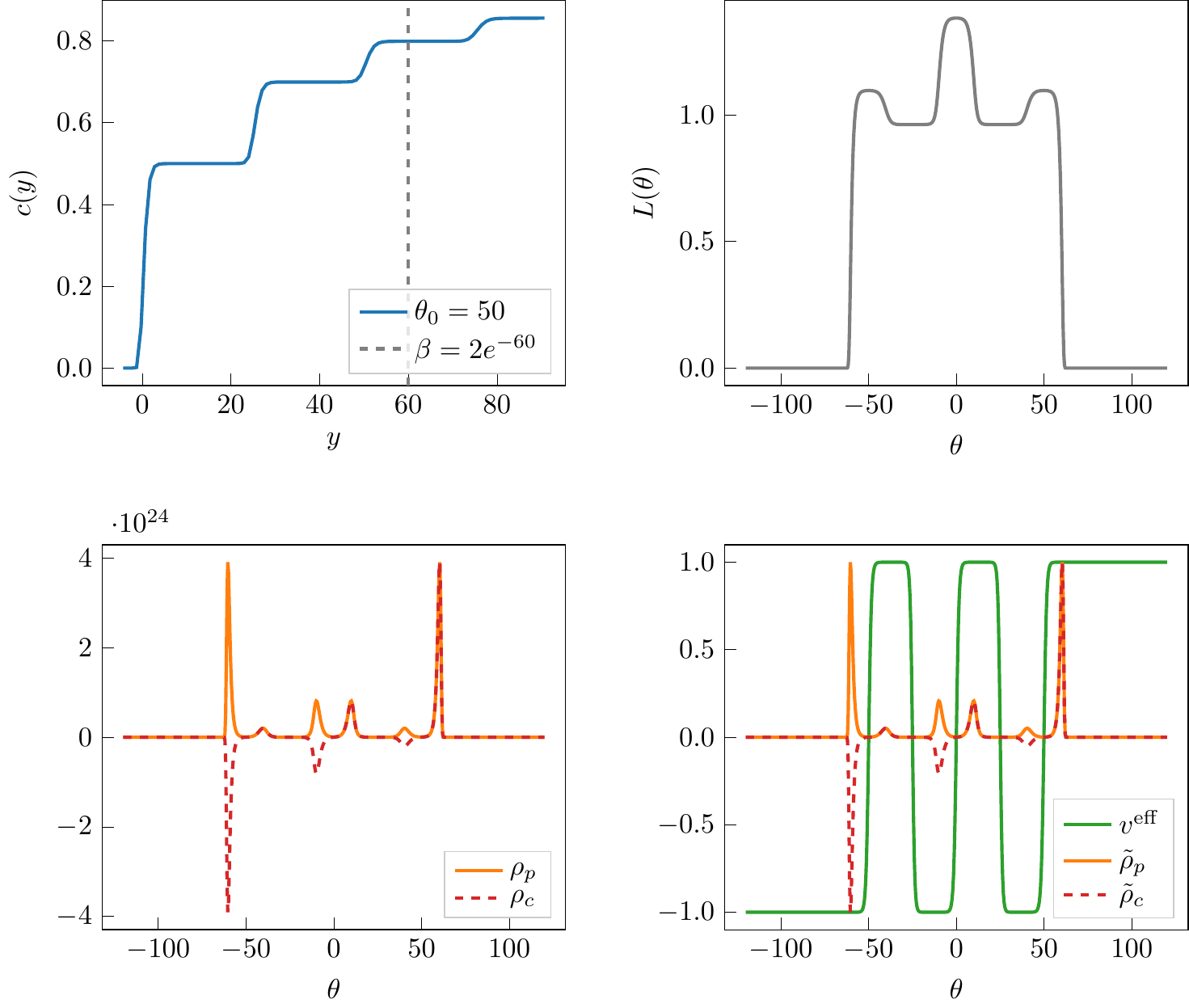}
	}
	\caption{Tetracritical Ising model for $k=3$: \emph{Top}: Scaling function and $L$-function for $\beta = 2e^{-60}$ and $\theta_0=50$, hence $1<\alpha = \frac{6}{5}<\frac{3}{2}$ \emph{Bottom Left}: Spectral density (solid orange line) and  spectral particle current (dashed red line). \emph{Bottom Right}: Effective velocity (green) versus scaled spectral particle functions (\ref{sca}). The effective velocity exhibits three plateaux with $v^{\text{eff}}(\theta)= 1$ and three plateaux with   $v^{\text{eff}}(\theta)= -1$. Therefore the effective velocity changes sign at $2k-1=5$ points.}\label{fig:collective_3}
\end{figure}

\begin{figure}
	\centering
		{
	\includegraphics{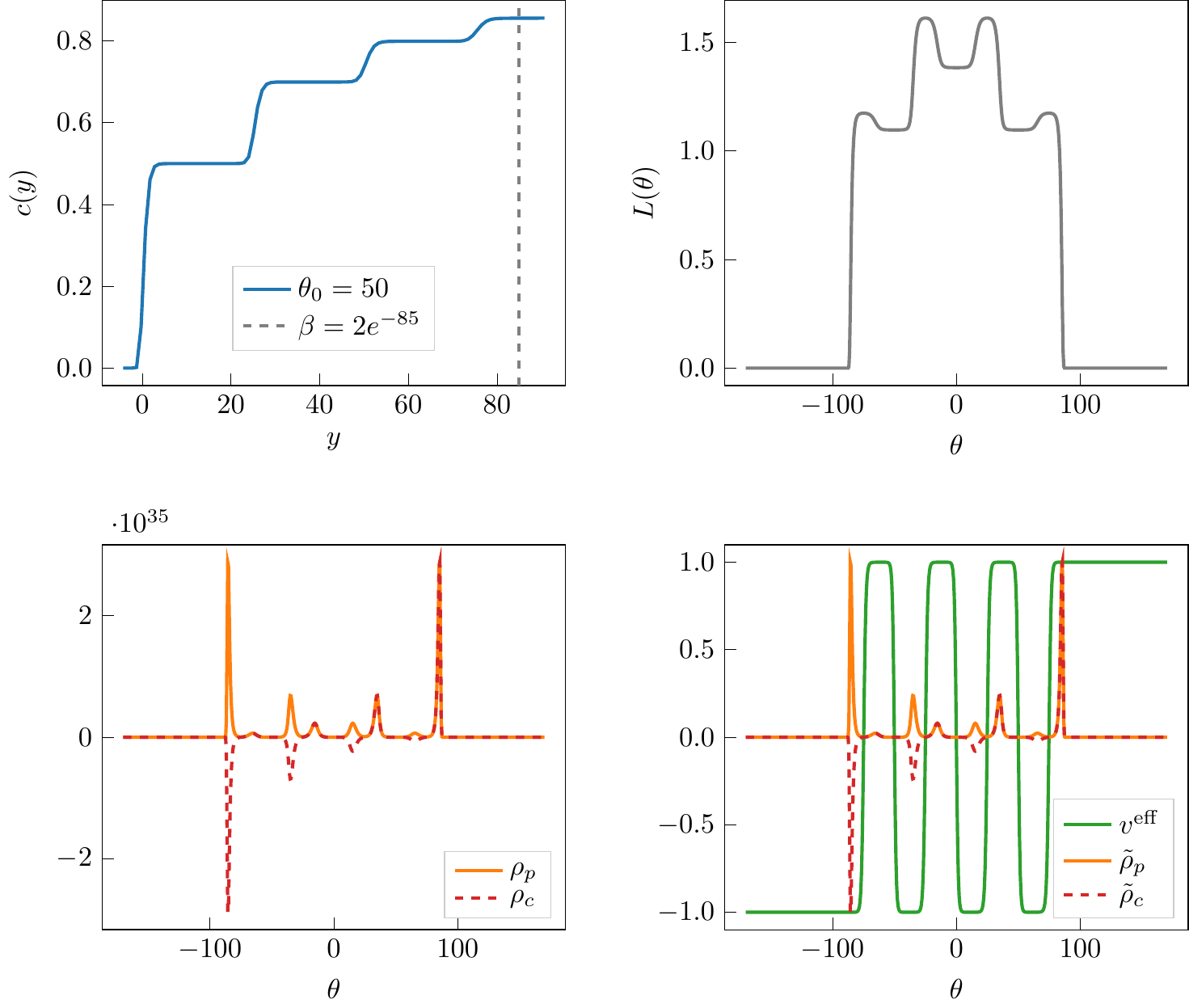}
	}
	\caption{Minimal model $\mathcal{M}_6$: \emph{Top}:Scaling function and $L$-function for  $\beta = 2e^{-85}$ and $\theta_0=50$, hence $\frac{3}{2}<\alpha = 1.7<2.$  \emph{Bottom Left}: Spectral density (solid orange line) and  spectral particle current (dashed red line). \emph{Bottom Right}: Effective velocity (green) versus scaled spectral particle functions (\ref{sca}). As in previous figures, we see that the effective velocity changes sign at $2k-1=7$ points and exhibits eight plateaux at alternating values $\pm 1$.}\label{fig:collective_4}
\end{figure}

Let us now solve the TBA-equation for the SM at some inverse temperature $\beta$, obtain $n(\theta)$ and compute from this the quantities (\ref{eq:charge}).  As expected, the form of these functions, similar to the $L$-functions,  will very much depend on $\alpha$. This is illustrated in Figs. \ref{fig:collective_1}, \ref{fig:collective_2}, \ref{fig:collective_3} and \ref{fig:collective_4}, all of which show the scaling function (highlighting one specific temperature) and the $L$-function, spectral particle density, spectral particle current and effective velocity for the same temperature. 

 Sitting on a step of the scaling function defined by an integer $k$ such that $\frac{k-1}{2}<\alpha<\frac{k}{2}$, the $L$-function has $2k$ kinks. At the same positions, the spectral particle density and spectral particle current exhibit  $2k$ bumps (local maxima).  Except for two outer-most bumps which are centered around $\pm y$, all other bumps are centered at values of $\theta$ whose distance from one of the outer-most bumps is a multiple of $\theta_0$. This is due to the structure of the scattering phase which maximizes interaction precisely for such distances. Thus, from a scattering viewpoint, we can think of each pair of peaks of the spectral particle density whose mutual distance is $\theta_0$ as describing densities of mutually interacting particles. In fact, it is this interaction that makes the particle density increase with respect to the non-interacting (free fermion) case shown in Fig.~\ref{fig:collective_1}.  A similar structure was found in \cite{ourunstable}. In all figures for this section we have introduced the convenient definitions:
 \beq
 \tilde{\rho}_{p/c}(\theta) = \frac{\rho_{p/c}(\theta)}{ \max{\rho_{p/c}(\theta)}}\,,
 \label{sca}
  \eeq
  which are the scaled spectral particle density ($p$) and current ($c$).
 
 A simple example of this phenomenology is provided in Fig.~\ref{fig:collective_2}. Here the two pairs of bumps centered at $\pm y$ and $\pm(y -\theta_0)$ describe densities of particles that are mutually interacting and also co-moving, as their effective velocities are the same. 
 
 If we actually compute the particle density associated to one of the larger maxima (say the right-most peak centered at $\theta=y$) and subtract from it the density associated with the corresponding peak of  a free fermion solution at the same temperature, we  find that the difference exactly matches the area of the smaller peak centered at $\theta=y-\theta_0$. We can say that the increase in the particle density with respect to the free theory is directly linked to the interaction with a smaller density of particles at distance $\theta_0$ in phase space. In other words, we can identify a stable density of interacting particles corresponding exactly to the area of the smaller peak. These particles are both interacting and co-moving as the two peaks of the spectral density also correspond to the same effective velocities ($+1$ in this example).

 The situation is similar but a bit more complicated in Figs.~\ref{fig:collective_3} and \ref{fig:collective_4}. For instance, in Fig.~\ref{fig:collective_3} the spectral particle density and current have six local maxima, which can be seen as describing three pairs of densities of mutually interacting particles centered around 
 $\pm y$, $\pm(y-\theta_0)$ and $\pm(y-2\theta_0)$. Again, interacting particles are associated with a fixed increase in particle density and are in all cases co-moving so that they can be easily identified by following the twists and turns of the effective velocity profile. Once more, the increase in the particle density can be attributed to interaction. This can be quantified by comparing the particle density associated to one of the larger maxima, say at $\theta=y$, with that of the same peak in the free fermion spectral particle density at the same temperature. The difference in areas coincides with the combined areas of peaks at $y-\theta_0$ and $y-2\theta_0$, both of which are mutually interacting with each other. Reflecting the interaction structure of the $\mathcal{M}A_3^{(+)}$ model, the peak at $y-\theta_0$ also interacts with the one at $y$.
 
 As $\alpha$ is further increased more pairs of these local maxima will emerge, until infinitely many are present in the deep UV limit. This behaviour is similar but distinct from what is observed in the $SU(3)_2$-HSG model studied in \cite{ourunstable}. The main difference arises from parity breaking in the HSG-model which for each particle type makes interaction maximal for some finite value of $\theta$ but not for its opposite. The effect of this lack of symmetry is that the spectral particle density of each particle never develops more than three local maxima. 
 
\begin{figure}[!ht]
	\centering
		{
		\includegraphics{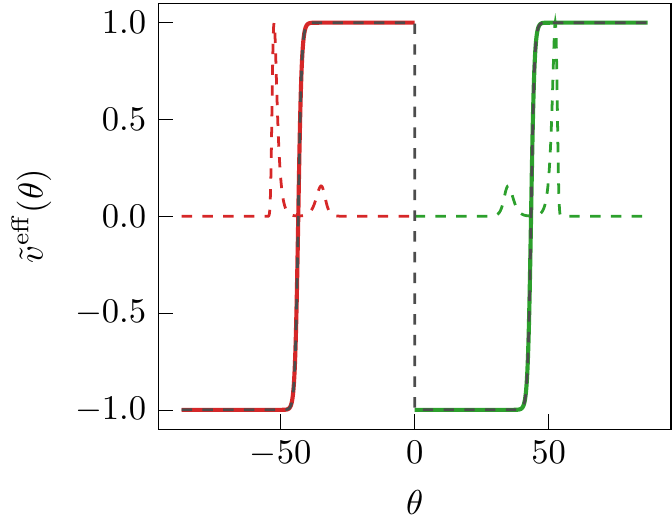}
		\includegraphics{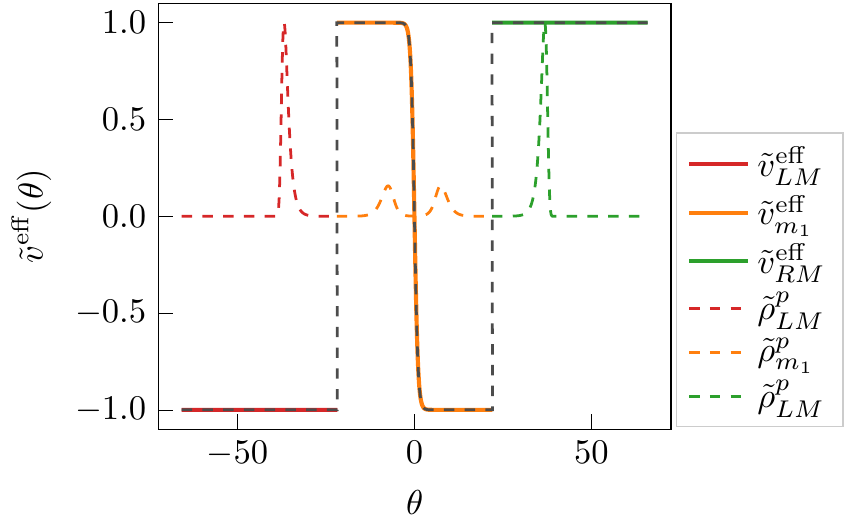}
		}
	\caption{Effective velocities and spectral particle density of the $\mathcal{M}A_2^{(+)}$ and $\mathcal{M}A_3^{(+)}$ models shifted according to the analysis in (\ref{eq:new_TBA})-(\ref{ctba}). \emph{Left}: UV limit of the $\mathcal{M}A_2^{(+)}$ model. We have two bumps of the spectral particle density for the RM and two bumps for the LM.  One peak is at $v^{\text{eff}}(\theta)=1$ (RM) and one at $v^{\text{eff}}(\theta)=-1$ (LM). \emph{Right}: IR limit of the $\mathcal{M}A_3^{(+)}$ model. LM (RM)  each have one peak in their spectral particle density at $v^{\text{eff}}(\theta)=\pm 1$, respectively.  The single magnon in the theory has two symmetric peaks (these are the two smaller peaks in each figure). The overall effective velocity  has the same features as the previous case and both descriptions match qualitatively the SM behaviour. The reconstructed functions reproduce all features of the functions presented in Fig.~\ref{fig:collective_2} for the case $k=2$.}\label{fig:k_2}
\end{figure}

\begin{figure}[!ht]
	\centering	
		{
		\includegraphics{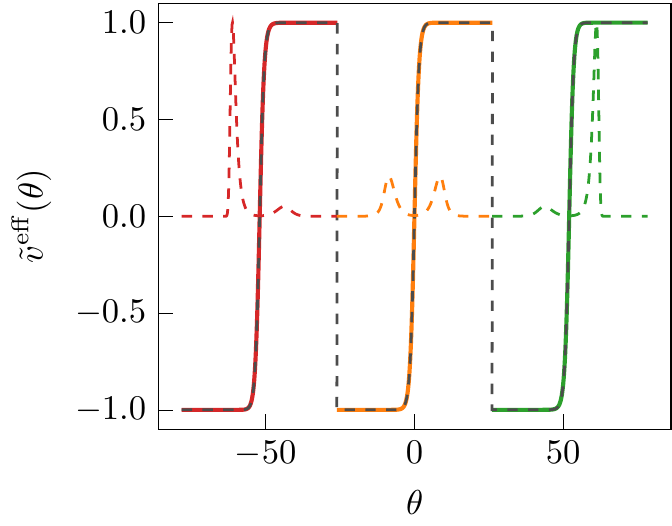}
		\includegraphics{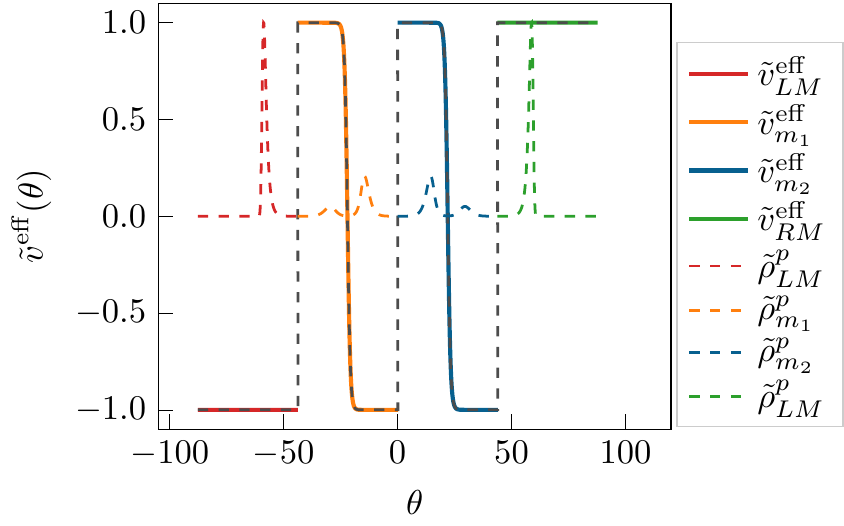}
		}
	
	\caption{Effective velocities and spectral particle density of the $\mathcal{M}A_3^{(+)}$ and $\mathcal{M}A_4^{(+)}$ models shifted according to the analysis in (\ref{eq:new_TBA})-(\ref{ctba}). \emph{Left}: UV limit of the $\mathcal{M}A_3^{(+)}$ model. We have two bumps of the spectral particle density for each excitation.  All peaks (for all excitations) are at $v^{\text{eff}}(\theta)=\pm 1$.  \emph{Right}: IR limit of the $\mathcal{M}A_4^{(+)}$ model. The reconstructed functions reproduce all features of the functions presented in Fig.~\ref{fig:collective_3} for the case $k=3$.}
 \label{fig:k_3}
\end{figure}

Unlike for the $L$-functions where a constant TBA analysis allows us to determine the exact height of each plateau, the definition of the effective velocities makes such an analysis difficult. However, 
 their complex structure can be well understood by  mapping the SM to the corresponding $\mathcal{M}A_k^{(+)}$ model described in the previous section. In particular we can reinterpret the multiple peaks of the spectral particle densities in terms of the LM, RM and magnonic  excitations  of the massless theory. This is shown in Fig. \ref{fig:k_2}  and \ref{fig:k_3} and is similar in spirit to the reconstruction of the $L$-function that we saw in Fig.~\ref{fig:limits}.

The structure of the effective velocity is far more peculiar. For $k>1$, $v^{\text{eff}}(\theta)$ becomes strongly non monotonic and develops $k$ plateaux at velocity $v^{\text{eff}}(\theta)=1$ and the same number at the opposite velocity. The $2k-1$ zeros of $v^\text{eff}(\theta)$ are at the midpoints of the $L$-function's internal plateaux, that is the points $z_i$ given by (\ref{eq:centers1}). This can be argued from the definitions \eqref{eq:TBAstair} and \eqref{dressing}, as these imply that
\begin{equation}
\label{eq:relation_dressing_derivative}
	\varepsilon'(\theta) = \beta \,p^{\text{dr}} (\theta)\,,
\end{equation}
and therefore the simple zeros of the effective velocity are at the points where $\varepsilon'(\theta)$ (and therefore $L'(\theta)$) vanishes. See Fig. \ref{fig:collective_1}, Fig. \ref{fig:collective_2}, Fig. \ref{fig:collective_3} and Fig. \ref{fig:collective_4} for some summarizing plots.

In Fig. \ref{fig:k_2}  and \ref{fig:k_3} we reconstruct the effective velocity of the SM for $k=2$ and $k=3$, respectively with the same procedure used for the $L$-function in section \ref{sec:mapping}. Notice that both the description as UV limit of the $\mathcal{M}A_2^{(+)}$ and IR limit of the $\mathcal{M}A_3^{(+)}$ model are valid. In the former case we have only two excitations with two peaks each in the spectral particle density; in the latter, LM and RM  have only one peak while the others are described by the magnons. Nonetheless, both scenarios are valid and all peaks are at the same values of the effective velocity.

How the intricate features of the spectral particle density, current and effective velocity change with temperature can be best seen in the video \cite{video} which provides a visual summary of all results in this Section.

\section{SM and the Partitioning Protocol}
In Section (\ref{GHDintro}) we introduced the main ideas behind the partitioning protocol.  In this section we focus on the average currents (\ref{j}) and densities (\ref{q}) for spins $s>1$. Unlike for the energy current and density studied in \cite{EFlow,CFTcur2, CFTcur}, it is not known precisely how higher spin currents and densities behave in the UV limit. The analysis below goes some way to explaining this and the SM model is an ideal theory to do so, as we are able to access multiple UV fixed points. 

\subsection{Higher Spin Currents and Densities}
Consider the one-particle eigenvalues (\ref{onep}) and their associated average current and density (\ref{q})-(\ref{j}) with $n(\theta)$ given by (\ref{npar}).
We are interested in the non-equilibrium steady state at ray $\xi=0$ and  in the situation in which the left and right subsystem both tend in the UV limit to the same minimal model, that is, defining
\beq
y_{\mathrm{R}}:=\log\frac{2}{\beta_{\mathrm{R}}}\,, \qquad y_{\mathrm{L}}:=\log\frac{2}{\beta_{\mathrm{L}}}\qquad \mathrm{and} \qquad \sigma:=\frac{\beta_{\mathrm{R}}}{\beta_{\mathrm{L}}}\,,
\eeq
we will consider
\begin{equation}
\frac{(k-1)\theta_0}{2}	< y_{\mathrm{R}} < y_{\mathrm{L}} <\frac{k\theta_0}{2},
\label{range}
\end{equation}
that is, $\sigma>1$. 
If $y_{\mathrm{L/R}}$ lie in the same range (\ref{range}) and $\sigma$ is not too large then the occupation functions of the right and left subsystems are very similar and one can numerically check that the zeroes of the effective velocity are still very close to their equilibrium values. Given (\ref{npar}) we can therefore define the total currents $\texttt{j}_{2s-1}$ (and $\texttt{j}_{2s}$) for the joint system as the sum of contributions from the two subsystems: 
\begin{equation}
	\texttt{j}_{2s-1} = \texttt{j}_{2s-1}^{R} + \texttt{j}_{2s-1}^{L},
\end{equation} 
with
\begin{equation}\label{eq:curr}
	 \texttt{j}_{2s-1}^{L/R} = \int_{L/R} \frac{d\theta}{2	\pi} p^{\mathrm{dr}} (\theta)n_{L/R}(\theta)h_{2s-1}(\theta)\,,
	\end{equation} 
and integration regions defined by
\begin{equation}
	\begin{aligned}
	R \equiv ]-\infty,z_1] & \cup [z_2,z_3] \cup \dots \cup [z_{2k-2}, z_{2k-1}]\\
	L \equiv [z_1, z_2] & \cup [z_3,z_4] \cup \dots \cup [z_{2k-1}, +\infty[.
\end{aligned}
\end{equation} 
Now we use the fact that if $\varepsilon_{L/R}(\theta)$ are the solutions of the TBA equations for the two Gibbs ensembles, from \eqref{eq:relation_dressing_derivative} it follows that
\begin{equation}
	\varepsilon'_{L/R}(\theta) = \beta_{L/R} \,p^{\text{dr}} (\theta).
\end{equation}
Therefore, since  $\varepsilon'_{L/R}(\theta) n_{L/R}(\theta) = - L'_{L/R}(\theta)$, we have
\begin{equation}\label{eq:expression}
		\texttt{j}_{2s-1}^{L/R}=-\frac{T_{L/R}^{s+1}}{2\pi}\int_{R}L'_{L/R}(\theta)\beta_{R/L}^{s}h_{2s-1}(\theta)\,.
\end{equation}
We will see in a few lines why this rewriting of $T_{L/R}=T_{L/R}^{s+1} \beta^s_{L/R}$ is useful. 
To proceed we can take $\theta_0, y_{R/L} \gg 1$ in order to exploit the correspondence between the SM and the $\mathcal{M}A_k^{(+)}$ model explained in the previous section. 
In this approximation the expression (\ref{eq:expression}) is slightly simpler, and is replaced by the massless limit

\begin{equation}
		{\texttt{j}}_{2s-1}^{R}= -\frac{T_{R}^{s+1}}{2^{2-s}\pi}\int_{K_1}d\theta\, e^{-s(y_{R}+\theta)}L'_{R}(\theta)\quad \mathrm{and} \quad 
		{\texttt{j}}_{2s-1}^{L}= -\frac{T_{L}^{s+1}}{2^{2-s}\pi}\int_{K_{2k}}d\theta\, e^{-s(y_{L}-\theta)}L'_{L}(\theta)\,.
\end{equation}
Integrating by parts and considering $y_{R,/L}$ sufficiently large, boundary terms vanish (as for large temperature, the $L$-functions exhibit a double-exponential decay) and we end up with 
\begin{equation}\label{eq:currents}
	\texttt{j}_{2s-1} = \frac{s}{2^{2-s}\pi} (\mathcal{C}^s_LT^{s+1}_L-\mathcal{C}^s_R T^{s+1}_R),
\end{equation}
where 
\begin{equation}\label{eq:coefficients}
	\mathcal{C}_R^s = \int_{K_{1}}d\theta e^{-s(y_R+\theta)} L_R(\theta), \hspace{1.5cm} 	\mathcal{C}_L^s = \int_{K_{2k}}d\theta e^{-s(y_L-\theta)} L_L(\theta).
\end{equation}
Since both the $L$-functions and the kinks $K_1$ and $K_{2k}$ are symmetric with respect to the origin we can write $\mathcal{C}_{L/R}^s:=\mathcal{C}^s(y_{L/R})$ with
	\begin{equation}\label{eq:C}
		\mathcal{C}^s(y)= \int_{K_{2k}}d\theta e^{-s(y-\theta)}L(\theta)\,.
	\end{equation}
Note that for $s=1$ and $(k-1)\theta_0/2<y<k\theta_0/2$, we recover the known result \cite{EFlow,CFTcur2, CFTcur}
\begin{equation}\label{eq:C_1}
	\lim\limits_{y,\theta_0 \to + \infty} \mathcal{C}^1(y) =\frac{\pi^2}{6}c_{k+2}\,,
\end{equation}
with $c_{k+2}$ given in (\ref{cp}). A very similar result can be obtained for $\texttt{j}_{2s}$ by simply replacing the minus sign by a plus sign in (\ref{eq:currents}).

\begin{figure}[!ht]
		{
		\includegraphics{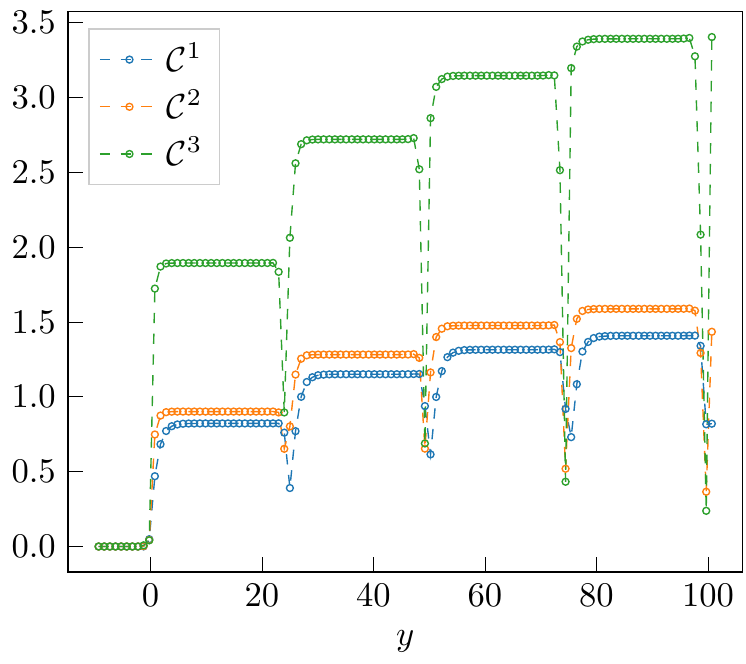}
		\includegraphics{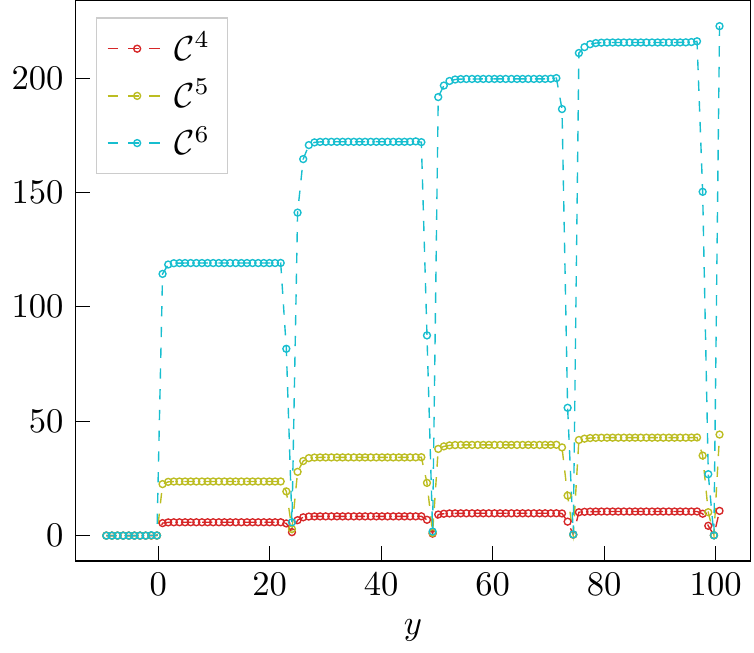}
		}
	\caption{$\mathcal{C}^s(y)$ for spin $s=1,2,3,4,5$ and 6 from Eq.~\eqref{eq:C}. Here $\theta_0=50$. } \label{fig:C_values}
\end{figure}
\begin{figure}
	\centering
    {
    \includegraphics{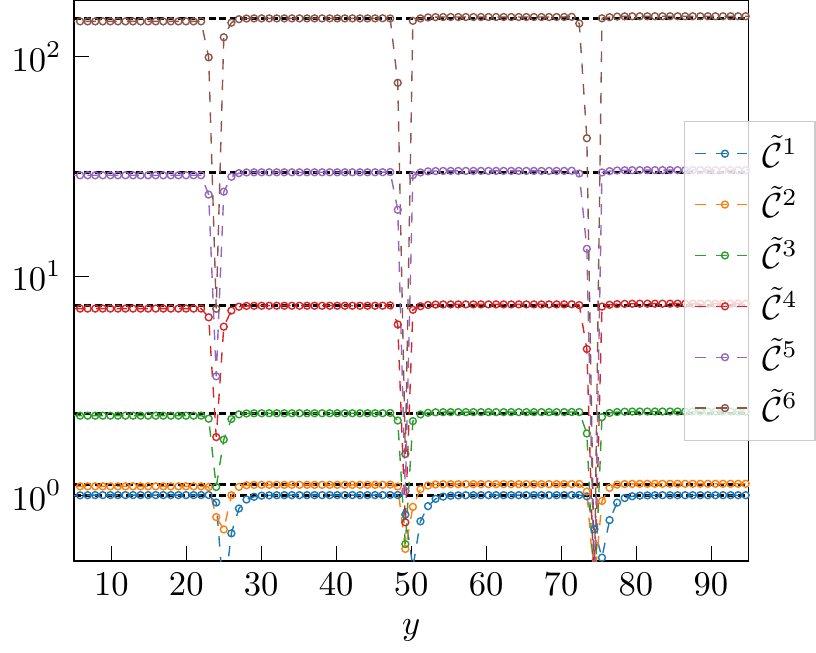}
    }    	
	\caption{ The function $\tilde{\mathcal{C}}^s(y)$ defined in (\ref{cscaled}) for spin $s=1,2,3,4,5$ and 6 from Eq.~\eqref{eq:C} and $\theta_0=50$. For each spin all plateaux are almost the same height, which strongly suggests the coefficient of the currents is very nearly proportional to the central charge $c_{k+2}$. However there are small deviations for spin $s>1$.}\label{fig:linearity}
\end{figure}
The same considerations can be applied to the charge densities $\texttt{q}_{2s-1}$ and $\texttt{q}_{2s}$  as defined in (\ref{q}). Again, the only contributions to the integrals come from $K_1$ and $K_{2k}$ and at a finite but large temperature one obtains:
\begin{equation}
	\texttt{q}_{2s-1} \approx \texttt{j}_{2s}\qquad\mathrm{and}\qquad	\texttt{q}_{2s} \approx \texttt{j}_{2s-1}.
\end{equation}
where the symbol $\approx$ indicates equality up to terms of order $O(e^{-y})$, thus becomes exact in the UV limit. Notice however that because of relativistic invariance $\texttt{q}_2 = \texttt{j}_1$ at any finite temperature.

Unfortunately, for $s>1$ the coefficients $\mathcal{C}^s(y)$ do not admit an obvious simple form. However we can at least perform some analysis based on numerics. Since it is only the value of the ratio $\alpha=\frac{y}{\theta_0}$ that fixes the UV limit of the theory, we expect that $\mathcal{C}^s(y_L) = \mathcal{C}^s(y_R)$ as long as the condition (\ref{range}) is satisfied. This is indeed verified by the numerical evaluation of the function $\mathcal{C}^s(y)$ which shows that it is constant in the central regions of $[(k-1)\theta_0/2, k \theta_0/2]$ (see Fig. \ref{fig:C_values}). 

\begin{figure}[!ht]
	\centering
		{
		\includegraphics{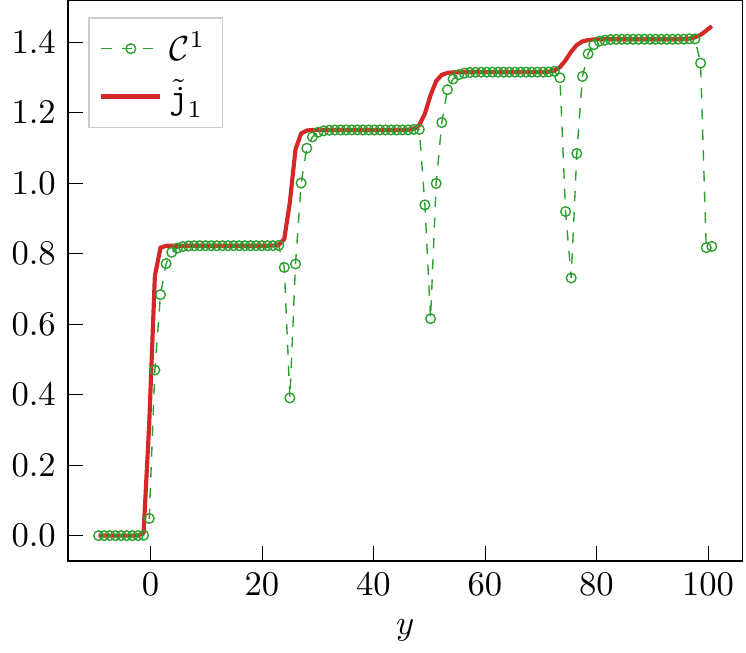}
		\includegraphics{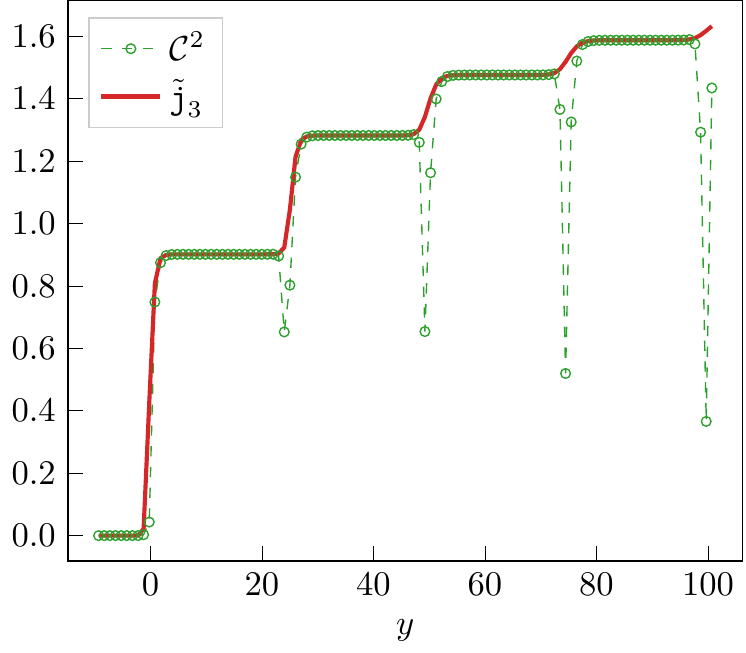}
		\includegraphics{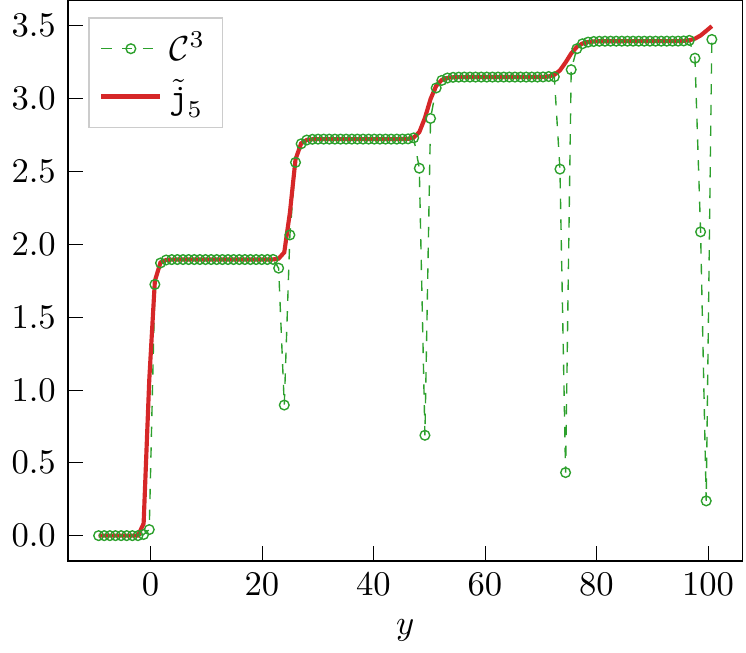}
		\includegraphics{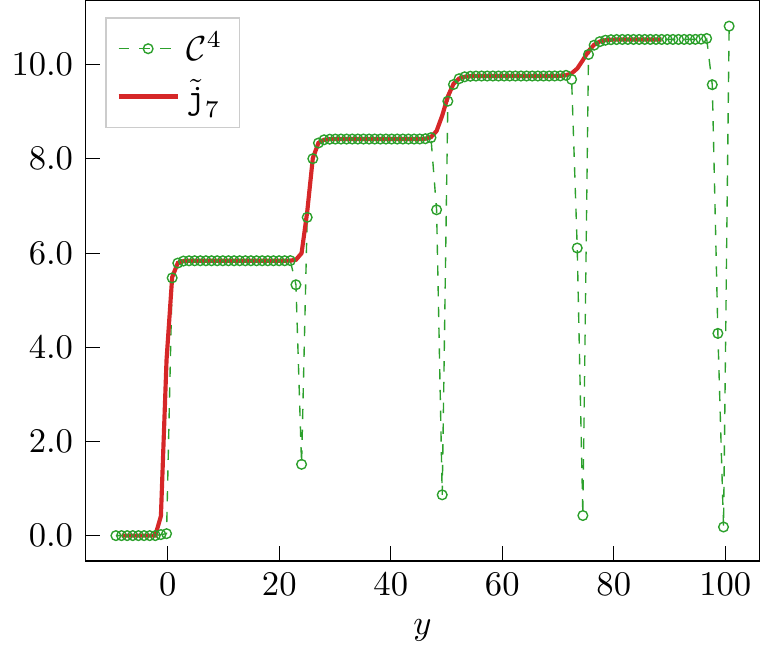}
		}
	\caption{Scaled currents $\tilde{\texttt{j}}_{2s-1}$ versus values of $\mathcal{C}^{s}(y)$ obtained from (\ref{eq:C}) for $s=1,2,3,4$. We see very good agreement in all plateau regions. 
	All data are for $\theta_0=50$ and $\sigma=e$.} \label{fig:currents}
\end{figure}

Looking at the plots in Fig. \ref{fig:C_values} we also observe that when $\alpha$ approaches a half-integer value the function $\mathcal{C}^s(y)$ has a sudden fall. 
This is consistent with the definition of $\mathcal{C}^s(y)$ in (\ref{eq:C}): it is easy to show that
\beq
 L(z_{2k}) \leq \lim\limits_{\alpha \to k/2} \mathcal{C}^s(y)\leq L(z_{2k-1})\,,
 \eeq
  where $L(z_{2k}) \simeq 0$  and  $L(z_{2k-1}) \simeq 1$.
In the central plateau regions, to a first approximation $\mathcal{C}^s(y)$ depends linearly on the central charge and exponentially on the spin. To see this we can treat $s$ as a continuous variable and differentiate $\mathcal{C}^s(y)$ with respect to $s$. We obtain a differential equation for $\mathcal{C}^s(y)$ which is solved with the initial condition (\ref{eq:C_1}):
\begin{equation}
	\mathcal{C}^s(y) = \frac{\pi^2}{6}c_{k+2} \exp{\int_1^s ds' \ev{\theta}_{s'} -(s-1)y},
\end{equation}
where
\begin{equation}
	\ev{\theta}_s \equiv \frac{\int_{K_{2k}}d\theta\theta e^{s\theta}L(\theta)}{\int_{K_{2k}}d\theta e^{s\theta}L(\theta)}.
\end{equation}
Writing $\mathcal{C}^s(y)$ in this form gives a complicated dependence on the spin but has the advantage that the central charge is factored out and corrections to linearity are encoded in the exponential.
In Fig. \ref{fig:linearity} we show the scaled functions 
\beq
\tilde{\mathcal{C}}^s_k(y) = \frac{6\,\mathcal{C}^s(y)}{\pi^2c_{k+2}}\,,
\label{cscaled}
\eeq
 for the first few integer values of $s$. As is clear from the figure, these functions have almost the same values at every plateau and deviations from linearity in $c_k$ are very small, but non-zero when $s>1$. 
 It would be interesting to investigate these corrections in more detail as well as the explicit dependence on the spin. The latter can be obtained exactly at the free fermion point, which is instructive. We show the calculation in Appendix~\ref{appendixb}.
 
 Another numerical check of (\ref{eq:C}) is presented in Fig.~\ref{fig:currents} where we look instead at the scaled currents:
 \begin{equation}
	\tilde{\texttt{j}}_{2s-1} = \frac{\pi\, \beta_R^{s+1}}{s\, 2^{s-2} (\sigma^{s+1}-1)} \texttt{j}_{2s-1}\,,
\end{equation}
as functions of $-y_L$ for fixed $\sigma=e$ along with the coefficients $\mathcal{C}^s(y)$ as obtained from (\ref{eq:C}). As expected, we find good agreement in all the plateau regions.

\section{Conclusions and Outlook}

In this paper we have used the generalized hydrodynamics approach to study the paradigmatic staircase model \cite{roaming}. The SM has a property which is unique among integrable quantum field theories, that is, at thermal equilibrium and high temperature, the theory does not flow to a single ultraviolet fixed point but rather to infinitely many, depending on how the high temperature limit is carried out. More precisely, at temperature $T$, the key scale in the model is the ratio $\alpha=\frac{y}{\theta_0}$ where $y=\log(2T)$ and $\theta_0$ is a parameter of the scattering matrix. Taking both $y$ and $\theta_0$ to infinity, whilst keeping their ratio fixed and finite, determines the choice of the $UV$ fixed point. The set of all possible $UV$ fixed points is given by the unitary minimal models of conformal field theory. In addition, the flow between consecutive CFTs, ordered according to their central charge, admits an effective description in terms of a massless theory known as the $\mathcal{M}A_k^{(+)}$ model, which we have also studied in this paper. 
The main conclusions of our study can be summarised as follows:

\begin{enumerate}
    \item The evaluation of TBA quantities that have a key hydrodynamic interpretation, such as the spectral particle current and density and the effective velocity gives new insights into the properties of the model, even at thermal equilibrium. The spectral functions of the SM generally exhibit multiple local maxima/minima while the effective velocity has multiple zeroes and plateaux. The number of such maxima, minima, zeroes and plateaux is linked to the scale $\alpha$ introduced above. For $\frac{k-1}{2}<\alpha<\frac{k}{2}$ the spectral particle density has $2k$ maxima and the spectral particle current has $k$ maxima and $k$ minima. The effective velocity has $2k-1$ zeroes,  $k$ plateaux at value $+1$ and $k$ plateaux at value $-1$. The non-monotonicity of the effective velocity is indeed one of the most distinct features of the model.
    All these properties can be seen in the video \cite{video}.
    \item The massless description of the SM by means of the $\mathcal{M}A_k^{(+)}$ models provides an accurate description of all the functions mentioned above. The agreement between both descriptions can be visualized by contrasting any of the SM functions with a ``cut and paste" arrangement of the corresponding functions found in the massless model. The agreement is particularly striking for the effective velocities, whose intricate square-wave structure is perfectly reproduced. 
    \item Similar properties are found within the partitioning protocol, especially when the right and left temperatures are chosen so as to fall within an interval of values of $\alpha$ associated to the same UV fixed point. In this context, the SM model provides an excellent opportunity to study the averages of conserved currents and densities for higher spins near different UV fixed points. Indeed, it is well-known how the energy current and density scale in CFT \cite{EFlow,CFTcur2,CFTcur} but a lot less is known for higher spin conserved quantities. In this paper we provide a partial answer to this question. We show numerically and analytically that higher spin currents and densities scale with appropriate powers of the temperatures $T_{L/R}$, as expected. More importantly, we also show that the numerical coefficient of these powers is very nearly proportional to the central charge of the UV fixed point. By very nearly we mean that the central charge is not exactly reproduced (except for spin 1, corresponding to the energy), but deviations from it are numerically very small and still to be analytically understood.   
\end{enumerate}

There are several future directions of research that we plan to pursue. First, in line with the last point, we would like to have a better analytical understanding of the UV scaling of the averages of higher spin currents and densities. We hope to achieve this through clever manipulation of the TBA equations (similar to how $c$ is obtained in terms of dilogarithm functions) and also through the explicit construction/study of conserved quantities in CFT. Second, we would like to generalize these results to other staircase models such as the generalizations constructed in \cite{DOREY_1993,GenDR}.

In conclusion, this paper constitutes an important addition to a so far small body of work including also \cite{ourunstable,tails} which explores how quantities of interest in the context of generalized hydrodynamics can provide, even at thermal equilibrium, new insights into properties of integrable quantum field theory. These insights are especially rich in IQFTs possessing unusual features, such as unstable particles or, as in the present case, an intricate RG physics. In all the models studied, including SM, we find a quantitative relationship between the emergence and area of new local maxima of the spectral particle density and the intensity of the interaction, as described by the two-particle scattering matrix. We also find that in such models the effective velocities can exhibit extremely intricate and unusual patterns as seen in \cite{video}, very far from the standard deformed $\tanh \theta$ shape that is found for most interacting IQFTs. 

\paragraph{Acknowledgments:}
We are grateful to C.~De Fazio, B.~Doyon, M.A.J.~Flores Herrera, E.~Guardati, M.~Mannatzu, and A.A.~Zi\'o\l{}kowska for useful discussions and especially to D.X. Horv\'ath for his assistance in clarifying some aspects of the numerical solution of the $\mathcal{M} A^{(+)}_k$ TBA equations. M.M., O.P. and F.R. thank the grant GAST of INFN and the fund RFO2021RAVA of Bologna University for partial financial support.

\appendix

\section{Constant TBA for the SM}
\subsection{$L$-function Kinks and Plateaux}
As mentioned in the introduction, the value of TBA functions in the SM depends very strongly on the interplay between the parameters $y=-\log(\beta/2)$ and $\theta_0$. In fact, the crucial parameter
is precisely their ratio
\begin{equation}
	\alpha = \frac{y}{\theta_0}
\end{equation}
If this ratio is kept fixed with both $y$ and $\theta_0$ going to infinity, then the scaling  function $c(y)$ tends to the central charge of one of the unitary minimal models $\mathcal{M}_{k+2}$.  
Precisely which one depends on the value $\alpha$.

 As already introduced in section \ref{sec:staircase}, the general structure is the following. When
\begin{equation}
	\frac{k-1}{2}<\alpha <\frac{k}{2}\hspace{0.3cm}, \hspace{1cm} k \in \mathbb{N},
\end{equation}
the $L$-function has $2k$ kinks at positions $\pm(y-\ell\theta_0)$ with $\ell=0,1,\dots,k-1$.  Kinks can be organized into two sequences, depending on whether they originate from the left or from the right. That is, we can order kinks according to
\begin{equation}
	\theta_{1,L} < 	\theta_{1,R} <	\theta_{2,L} < \dots < \theta_{k,L} < \theta_{k,R}
\end{equation}
with
\begin{equation}
	\begin{cases}
	\theta_{\ell,L} = -y + (\ell-1)\theta_{0}\\
		\theta_{\ell,R} = y - (k-\ell)\theta_{0}
			\end{cases}
	\hspace{0.5cm}	\ell = 1,2,\dots,k
\end{equation}
As we have seen, the $L$-function will develop plateaux in the regions between consecutive kinks. There are therefore $2k+1$ plateaux, the first and the last being in the regions $\theta < -y$ and $\theta > y$, respectively, where $L(\theta)$ is effectively zero. The internal plateaux are of alternating width, and so is the distance between two adjacent kinks. We arrange the internal plateaux in two sequences, odd and even, according to:

\begin{equation}
	\begin{cases}
	P_{2\ell-1} = [\theta_{\ell,L}, \theta_{\ell,R}], \hspace{0.5cm} \ell = 1,\dots,k\\
	P_{2\ell} = [\theta_{\ell,R}, \theta_{\ell+1,L}], \hspace{0.5cm} \ell = 1,\dots,k-1,
	\end{cases}
\end{equation}
with $|P_{2\ell-1}| = 2y-(k-1)\theta_0$ and $|P_{2\ell}| = k\theta_0-2y$, where $|P|$ indicates the width of plateau $P$.
\begin{figure}[h!]
	\centering
	\begin{tikzpicture}
	\node at (0.05,0.05) {
	\includegraphics{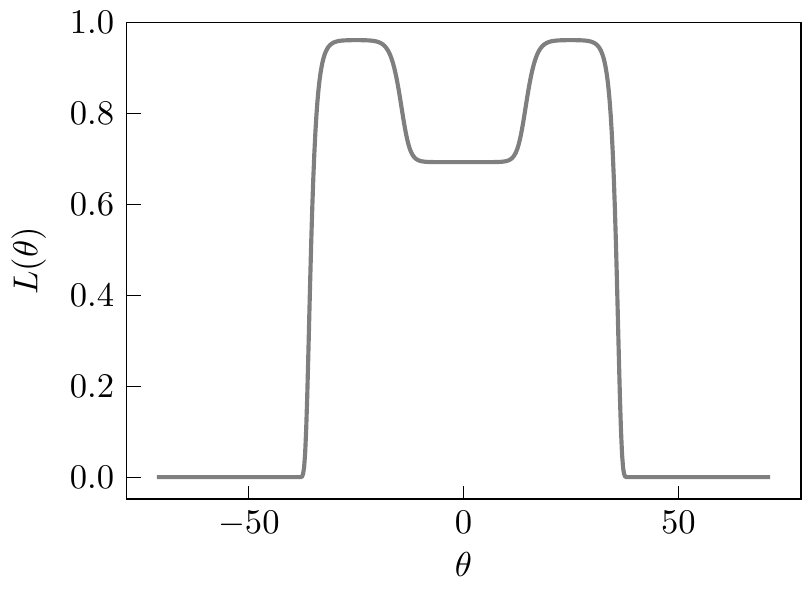}
	};
		\node[] at (-1.5, -1.5) {$z_0$};
		\node[] at (2.85, -1.5) {$z_4$};
		\node[] at (-0.4, 2.3) {$z_1$};
		\node[] at (1.8, 2.3) {$z_3$};
		\node[] at (0.68, 1.1) {$z_2$};
		
		\node[] at (-1.2, 0.5) {\footnotesize{$K_1$}};
		\node[] at (0.35, 1.9) {\footnotesize{$K_2$}};
		\node[] at (1, 1.9) {\footnotesize{$K_3$}};
		\node[] at (2.6, 0.5) {\footnotesize{$K_4$}};
		
		\draw[] (-1.5, -1.8) [fill=red] circle(0.07);
		\draw[] (2.85, -1.8) [fill=red] circle(0.07);
		
		\draw[] (-0.4, 2.63) [fill=red] circle(0.07);
		\draw[] (1.80, 2.63) [fill=red] circle(0.07);
		
		\draw[] (0.68, 1.4) [fill=red] circle(0.07);
		
	\end{tikzpicture}
	\caption{$L$-function of the SM for $k=2$. Points $z_\ell$ lie at the centre of plateaux, while kinks $K_\ell$ interpolate between them. }\label{fig:notation}
\end{figure}
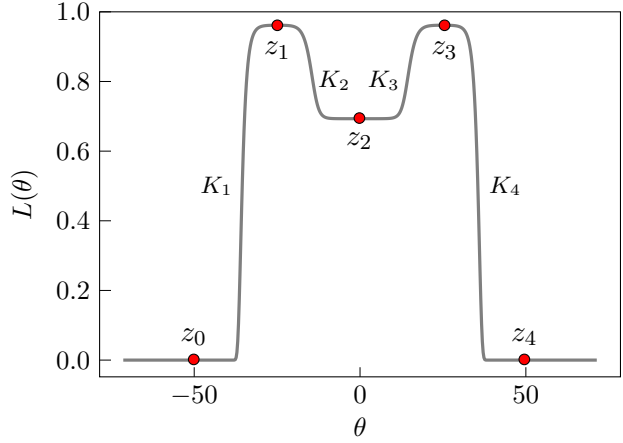

Finally, we define the mid-points of the internal plateaux
\begin{equation}
	z_\ell \equiv \frac{(\ell-k)\theta_0}{2}, \hspace{1cm} 1,\dots,2k-1, \label{eq:centers1}
\end{equation}
and
\begin{equation}
	z_0 \equiv -\frac{k\theta_0}{2},\hspace{0.5cm}	z_{2k} \equiv \frac{k\theta_0}{2} \label{eq:centers2}
	\end{equation}
to extend the definition (\ref{eq:centers1}) to the first and last plateaux where $L(\theta)$ is zero. All these definitions are illustrated in Fig.~\ref{fig:notation} for the case $k=2$.
So far we have ignored the fact that the kinks $K_i$ have themselves a finite width. We can see that the $\ell^{\text{th}}$ kink is spread in the interval
\begin{equation}
	K_\ell = [z_{\ell-1},z_\ell], \hspace{0.5cm}i=1,2,\dots, 2k\,.
\end{equation}

\subsection{Constant TBA equations}\label{sec:mapping}
Following the definitions given in the previous section, we will now proceed to derive the so-called constant TBA equations \cite{tba1,tba12,tba2}. 
These can be seen as UV limits of the original TBA  equation for the SM, whose solutions describe the height of the centres of the plateaux at $z_i$. Because there are many such plateaux in the SM there are several constant TBA equations, which in effect form a system of coupled equations in much the same spirit as the TBA description of the $\mathcal{M}A_k^{(+)}$ model presented in section~\ref{Anseries}. 
Following the steps used in \cite{DOREY_1993,GenDR}, we want to look at the values of $L(\theta)$ at the midpoints of the plateaux. Let $\varepsilon_i:=\varepsilon(z_i)$ be the values of the pseudoenergy at the plateaux' mid-points.  The TBA-equation can be expressed as
\begin{equation}
	\varepsilon_i = 2e^{-y}\cosh(z_i)-\sum_{i=1}^{2k}\int_{K_i} \frac{d\theta'}{2\pi} L(z_i-\theta')\varphi(\theta')\,,
	\label{conv}
\end{equation}
where we can reduce our integration region because of the double-exponential fall-off of $L(\theta)$.
The driving term is vanishingly small for all $i$ except for regions $K_1$ and $K_{2k}$:
\begin{equation}\label{eq:dr}
	2e^{-y}\cosh(\theta) \sim 
	\begin{cases}
		\begin{aligned}
		e^{-y-\theta}, \hspace{1cm}&\theta \in K_1\\
		0,\hspace{1.5cm}&\theta \in K_2,\dots,K_{2k-1}\\
		e^{-y+\theta},\hspace{1cm}&\theta \in K_{2k}
		\end{aligned}
	\end{cases}
\end{equation}
Furthermore the kernel couples only those mid-points which are at distance $\theta_0$ from each other since it is strongly peaked at $\theta' = \pm \theta_0$. Therefore, we can identify  $L(z_i\pm\theta_0) := L_{i\pm2}$. We can also make the standard asumption that underlies all constant TBA derivations, namely that the $L$-function is effectively constant (with values $L_{i\pm 2}$) over the region where the kernel is non-vanishing. In effect this means that we can take the $L$-functions outside the convolution (\ref{conv}) and just integrate each term in the kernel separately, giving
\beq
\int_{-\infty}^\infty \frac{d\theta}{\cosh(\theta\pm\theta_0)} =\pi \,.
\label{eq:new_TBA}
\eeq
The equations for plateaux midpoints simplify to:
\begin{equation}
	-2\varepsilon_i = L_{i+2}+L_{i-2}, \hspace{0.5cm} i=1,\dots,2k-1\,,
	\label{ctba}
\end{equation}
with boundary conditions $\varepsilon_{-1} = \varepsilon_{0} = \varepsilon_{2k} = \varepsilon_{2k+1} = \infty$. Setting $x_i = \exp{-\varepsilon_{2i-1}}$  for $i=1,\dots,k$ and  $y_i = \exp{-\varepsilon_{2i-2}}$ for $i=2,\dots,k$ we
can rewrite (\ref{ctba}) as
\begin{equation}\label{eq:sets}	
		x_i^2 = \prod_{j=1}^k (1+x_{j})^{I_{ij}}\qquad {\mathrm{and}} \qquad
		y_i^2 =\prod_{j=2}^k(1+y_{j})^{I_{ij}}\,,
\end{equation}
with boundary conditions $x_0 = x_{k+1} = y_1 = y_{k+1} = 0$. These two sets of equations are again reminiscent of an underlying $A_n$ structure and $I_{ij}$ is the incidence matrix of $A_n$.  Solutions to these equations where first given in \cite{elastic} in their study of the constant TBA of $A_n$ minimal Toda field theory and also in 
	 \cite{ZAMOLODCHIKOV1991497} in the study of RSOS models. They were also found in a more general context in \cite{kuniba}
	\beq
	x_\ell=\frac{\sin\frac{\pi\ell }{k+3} \sin \frac{\pi(\ell+2) }{k+3}}{\sin^2\frac{\pi }{k+3} }, \qquad y_\ell=\frac{\sin\frac{\pi\ell }{k+2} \sin \frac{\pi(\ell+1) }{k+2}}{\sin^2\frac{\pi }{k+2}}\,.
	\eeq
	The associated UV central charge can be computed in terms of these values in the standard way \cite{tba1,tba12} and gives $c_{k+2}$ as in equation (\ref{cp}). 
	
	Due to the kink structure of the $L$-function discussed here (see (\ref{eq:dr})) the SM scaling function (\ref{Ecy})  only receives contributions from the outer-most kinks $K_1$ and $K_{2k}$. 
	Writing $c(y,\theta_0) = c_{-}+c_{+}$ with
\begin{equation}
	c_{-} = \frac{3}{\pi^2}\int_{K_1} d \theta e^{-y-\theta}L(\theta), \hspace{0.5cm}c_{+} = \frac{3}{\pi^2}\int_{K_{2k}} d \theta e^{-y+\theta}L(\theta)\,,
\end{equation}
we find that $c_{-} = c_{+}$ and  the two contributions match exactly the formula (\ref{scaling}) for the $\mathcal{M}A_k^{(+)}$ model. Their computation is standard and, as expected, for $\frac{k-1}{2}<\alpha<\frac{k}{2}$ we obtain once more $c_{k+2}$ as in (\ref{cp}).

\section{Higher Spin Currents  for a Free Fermion}
\label{appendixb}

\begin{figure}[!ht]
	\centering
	{\includegraphics[width=7.6cm]{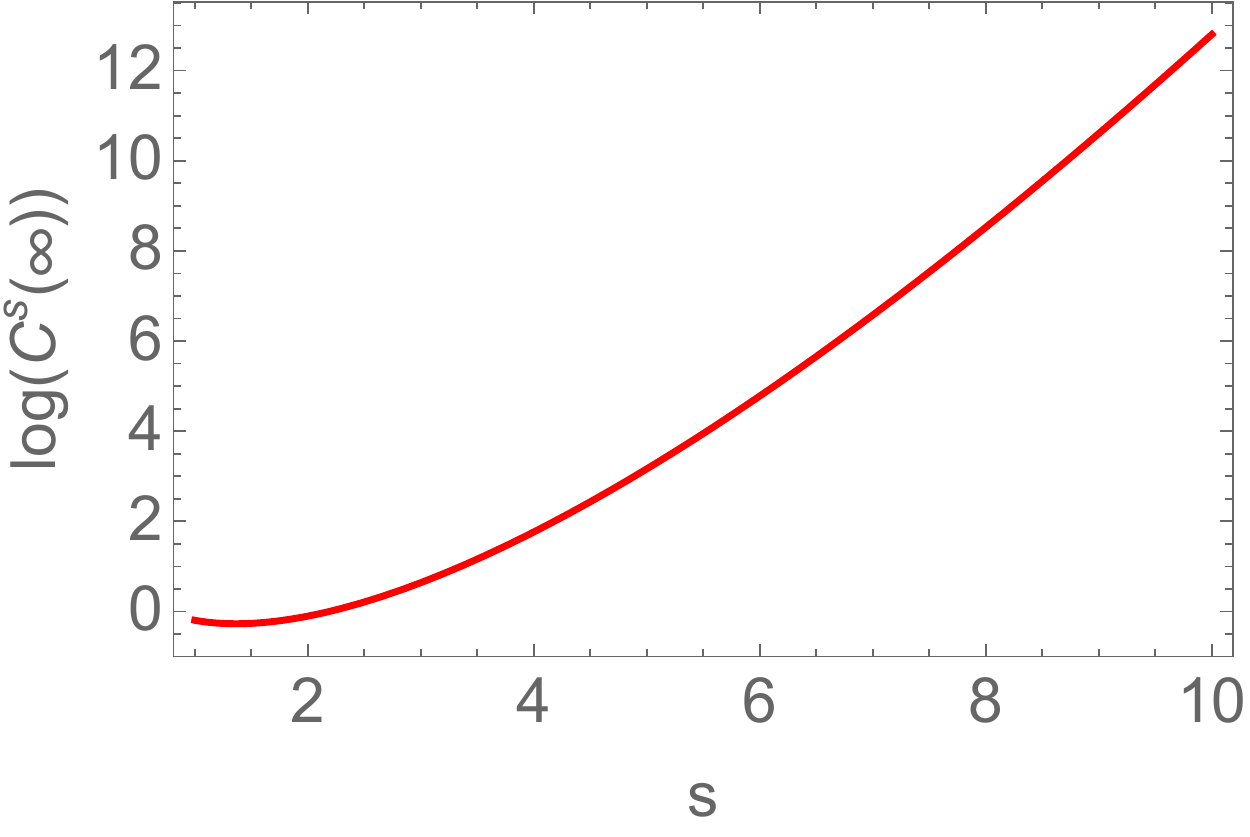}}
	\caption{$\log(\mathcal{C}^s(\infty))$ for the free fermion.} \label{cs}
\end{figure}

In the free fermion case we can evaluate explicitly and analytically all the functions that we have been discussing in this paper. In particular, in the partitioning protocol we have that
\beq
\varepsilon_{L/R}(\theta)=\beta_{L/R} \cosh\theta\,,
\eeq
and so the even and odd spin currents can be written as
\beqa
\texttt{j}_{2s-1}&=&\frac{1}{2\pi} \int_{0}^\infty \cosh(s \theta) \sinh\theta \left[\frac{1}{1+e^{\beta_L \cosh\theta}}-\frac{1}{1+ e^{\beta_R \cosh \theta}} \right]\,,\\
\texttt{j}_{2s}&=&\frac{1}{2\pi} \int_{0}^\infty \sinh(s \theta) \sinh\theta \left[\frac{1}{1+e^{\beta_L \cosh\theta}}+\frac{1}{1+ e^{\beta_R \cosh \theta}} \right]\,.
\eeqa
For $\beta > 0$ it is possible to expand the occupation functions as
\beq
\frac{1}{1+e^{\beta \cosh\theta}}=e^{-\beta \cosh\theta} \sum_{n=0}^\infty (-1)^n {e^{-n \beta \cosh\theta}}\,,
\eeq
and to express the CFT limit of the currents above in terms of modified Bessel functions $K_s(x)$.
For instance, for the currents $\texttt{j}_{2s}$ we can use the identity
\beq
\int_0^\infty \sinh(s \theta) \sinh \theta  e^{-\beta \cosh \theta}\, d\theta=\frac{s}{\beta}K_s(\beta)\,,
\eeq
to obtain
\beqa
\texttt{j}_{2s}&=&\frac{1}{2\pi} \sum_{n=0}^\infty (-1)^n \int_{0}^\infty \sinh(s \theta) \sinh\theta [e^{-\beta_L (n+1) \cosh\theta}+e^{-\beta_R (n+1) \cosh\theta}] d\theta\nonumber\\
&=& \frac{s}{2\pi} \sum_{n=0}^\infty \frac{(-1)^n}{n+1} \left[T_L K_s((n+1)\beta_L)+T_R K_s((n+1)\beta_R)\right]\nonumber\\
&\approx & \frac{s}{2\pi} \sum_{n=0}^\infty \frac{(-1)^n}{(n+1)}\left[
\frac{2^{s-1}\Gamma(s)}{ \beta_L^{s+1} (n+1)^s}- \frac{2^{s-1}\Gamma(s)}{\beta_R^{s+1} (n+1)^s}\right]\nonumber\\
&=&  \frac{s \Gamma(s)}{2^{2-s} \pi} (T_L^{s+1}+T_R^{s+1})\sum_{n=0}^\infty \frac{(-1)^n}{(n+1)^{s+1}}\nonumber\\
&=&  \frac{s \Gamma(s)}{4 \pi} (2^s-1) \zeta(s+1) (T_L^{s+1}+T_R^{s+1})\,,
\eeqa
where we have used the small $x$ expansion of $K_s(x)$ and $\zeta(x)$ is Riemann's zeta function. As discussed in the main text, we expect the coefficients (\ref{eq:C}) to be identical for $\texttt{j}_{2s-1}$ and $\texttt{j}_{2s}$  and so from the computation above we can identify
\beq
\mathcal{C}^s(\infty) = \Gamma(s) \zeta(s+1) (1-2^{-s})\,.
\eeq
In particular, we have the values
\beqa
&& \mathcal{C}^1(\infty)=\frac{\pi^2}{12}=0.822467\,, \quad \mathcal{C}^2(\infty)=\frac{3\zeta(3)}{4}=0.901543\,, \nonumber\\
&&  \mathcal{C}^3(\infty)=\frac{7\pi^4}{360}=1.89407\,,  \quad \mathcal{C}^4(\infty)=\frac{45\zeta(5)}{8}=5.83272\,.
\eeqa
These values are well reproduced by the height of the first plateau in the Figs.~\ref{fig:C_values} and \ref{fig:currents}. 
As we can see in Fig.~\ref{cs} this coefficient grows rapidly with the spin $s$. 

In fact, using Stirling's approximation we have that
\beq
\log(\mathcal{C}^{s+1}(\infty)) \approx s\log s-s.
\eeq


\clearpage
\begin{thebibliography}{10}

\bibitem{20}
R.~Zwanzig,
\newblock {Non-equilibrium Statistical Physics},
\newblock New York, Oxford University Press  (2001).

\bibitem{Eisert}
J. Eisert, M. Friesdorf and C. Gogolin, Quantum many-body systems out of equilibrium,
Nature Phys. 11 (2015) 124.

\bibitem{kinoshita}
T.~Kinoshita, T.~Wenger, and D.~Weiss,
\newblock {A Quantum Newton's Cradle},
\newblock Nature {\bf 440}, 900 (2006).

\bibitem{Rigol}
M.~Rigol, V.~Dunjko, V.~Yurovsky, and M.~Olshanii,
\newblock Relaxation in a Completely Integrable Many-Body Quantum System: An Ab
  Initio Study of the Dynamics of the Highly Excited States of 1D Lattice
  Hard-Core Bosons,
\newblock Phys. Rev. Lett. {\bf 98}, 050405 (2007).

\bibitem{NERev2}
F.~H.~L. Essler and M.~Fagotti,
\newblock Quench dynamics and relaxation in isolated integrable quantum spin
  chains,
\newblock J. Stat. Mech. {\bf 2016}(6), 064002 (2016).

\bibitem{NERev3}
R.~Vasseur and J.~E. Moore,
\newblock Nonequilibrium quantum dynamics and transport: from integrability to
  many-body localization,
\newblock J. Stat. Mech. {\bf 2016}(6), 064010 (2016).

\bibitem{NERev4}
E.~Ilievski, M.~Medenjak, T.~Prosen, and L.~Zadnik,
\newblock Quasilocal charges in integrable lattice systems,
\newblock J. Stat. Mech. {\bf 2016}(6), 064008 (2016).

\bibitem{NERev5}
D.~Bernard and B.~Doyon,
\newblock Conformal field theory out of equilibrium: a review,
\newblock J. Stat. Mech. {\bf 2016}(6), 064005 (2016).

\bibitem{NERev6}
L.~D'Alessio, Y.~Kafri, A.~Polkovnikov, and M.~Rigol,
\newblock From quantum chaos and eigenstate thermalization to statistical
  mechanics and thermodynamics,
\newblock Adv. in Phys. {\bf 65}(3), 239--362 (2016).

\bibitem{NERev7}
C.~Gogolin and J.~Eisert,
\newblock Equilibration, thermalisation, and the emergence of statistical
  mechanics in closed quantum systems,
\newblock Rep. on Prog. in Phys. {\bf 79}(5), 056001 (2016).

\bibitem{Brev}
B.~Doyon,
\newblock Lecture notes on Generalised Hydrodynamics,
\newblock SciPost Phys. Lecture Notes  (2020).

\bibitem{CEM}
P.~Calabrese, H.~Essler, and G.~Mussardo~(ed.),
\newblock {Quantum Integrability in Out-of-Equilibrium Systems},
\newblock J. Stat. Phys. 064001 (2016).

\bibitem{ourhydro}
O.~A. Castro-Alvaredo, B.~Doyon, and T.~Yoshimura,
\newblock {Emergent hydrodynamics in integrable quantum systems out of
  equilibrium},
\newblock Phys. Rev. {\bf X6}(4), 041065 (2016).

\bibitem{theirhydro}
B.~Bertini, M.~Collura, J.~De~Nardis, and M.~Fagotti,
\newblock {Transport in Out-of-Equilibrium $XXZ$ Chains: Exact Profiles of
  Charges and Currents},
\newblock Phys. Rev. Lett. {\bf 117}(20), 207201 (2016).

\bibitem{cur1}
M.~Fagotti,
\newblock Charges and currents in quantum spin chains: late-time dynamics and
  spontaneous currents,
\newblock J. Phys. {\bf A50}(3), 034005 (2016).

\bibitem{cur2}
A.~Urichuk, Y.~Oez, A.~Kl\"umper, and J.~Sirker,
\newblock The spin Drude weight of the XXZ chain and generalized hydrodynamics,
\newblock SciPost Phys. {\bf 6}(1) (2019).

\bibitem{cur3}
D.-L. Vu and T.~Yoshimura,
\newblock Equations of state in generalized hydrodynamics,
\newblock SciPost Phys. {\bf 6}(2) (2019).

\bibitem{dNBD2}
J.~De~Nardis, D.~Bernard, and B.~Doyon,
\newblock Diffusion in generalized hydrodynamics and quasiparticle scattering,
\newblock SciPost Phys. {\bf 6}(4) (2019).

\bibitem{cur4}
Z.~Bajnok and I.~Vona,
\newblock Exact finite volume expectation values of conserved currents,
\newblock Phys. Lett. {\bf B805}, 135446 (2020).

\bibitem{cur5}
H.~Spohn,
\newblock Collision rate ansatz for the classical Toda lattice,
\newblock Phys. Rev. {\bf 101}(E6) (2020).

\bibitem{cur6}
T.~Yoshimura and H.~Spohn,
\newblock Collision rate ansatz for quantum integrable systems,
\newblock SciPost Phys. {\bf 9}(3) (2020).

\bibitem{cur7}
M.~Borsi, B.~Pozsgay, and L.~Pristy\'ak,
\newblock Current Operators in Bethe Ansatz and Generalized Hydrodynamics: An
  Exact Quantum-Classical Correspondence,
\newblock Phys. Rev. {\bf 10}(X1) (2020).

\bibitem{cur8}
B.~Pozsgay,
\newblock Current operators in integrable spin chains: lessons from long range
  deformations,
\newblock SciPost Phys. {\bf 8}(2) (2020).

\bibitem{cur9}
B.~Pozsgay,
\newblock Algebraic Construction of Current Operators in Integrable Spin
  Chains,
\newblock Phys. Rev. Lett. {\bf 125}(7) (2020).

\bibitem{tba1}
A.~Zamolodchikov,
\newblock {Thermodynamic Bethe ansatz in relativistic models. Scaling three
  state Potts and Lee-Yang models},
\newblock Nucl. Phys. {\bf B342}, 695--720 (1990).

\bibitem{tba12} T.~R.~Klassen and E.~Melzer, Purely elastic scattering theories and their ultraviolet limits, Nucl. Phys. {\bf B338} 485--528 (1990).

\bibitem{tba2}
T.~R.~Klassen and E.~Melzer,
\newblock {The Thermodynamics of purely elastic scattering theories and
  conformal perturbation theory},
\newblock Nucl. Phys. {\bf B350}, 635--689 (1991).

\bibitem{EFlow}
D.~Bernard and B.~Doyon,
\newblock Energy flow in non-equilibrium conformal field theory,
\newblock J. Phys. {\bf 45}(A36), 362001 (2012).

\bibitem{CFTcur2}
D.~Bernard and B.~Doyon,
\newblock Time-reversal symmetry and fluctuation relations in non-equilibrium
  quantum steady states,
\newblock J. Phys. {\bf A46}(37), 372001 (2013).

\bibitem{CFTcur}
D.~Bernard and B.~Doyon,
\newblock Non-Equilibrium Steady States in Conformal Field Theory,
\newblock Annales Henri Poincar\'e {\bf 16}(1), 113--161 (2014).

\bibitem{DY}
B.~Doyon and T.~Yoshimura,
\newblock A note on generalized hydrodynamics: inhomogeneous fields and other
  concepts,
\newblock SciPost Phys. {\bf 2}(2) (2017).

\bibitem{bastianello2019Integrability}
A.~Bastianello and A.~De~Luca,
\newblock Integrability-Protected Adiabatic Reversibility in Quantum Spin
  Chains,
\newblock Phys. Rev. Lett. {\bf 122}(24) (2019).

\bibitem{BasGeneralised2019}
A.~Bastianello, V.~Alba, and J.-S. Caux,
\newblock Generalized Hydrodynamics with Space-Time Inhomogeneous Interactions,
\newblock Phys. Rev. Lett. {\bf 123}(13) (2019).

\bibitem{dNBD}
J.~De~Nardis, D.~Bernard, and B.~Doyon,
\newblock Hydrodynamic Diffusion in Integrable Systems,
\newblock Phys. Rev. Lett. {\bf 121}(16) (2018).

\bibitem{GHKV2018}
S.~Gopalakrishnan, D.~A. Huse, V.~Khemani, and R.~Vasseur,
\newblock Hydrodynamics of operator spreading and quasiparticle diffusion in
  interacting integrable systems,
\newblock Phys. Rev. {\bf B98}(22) (2018).

\bibitem{FagottiLocally}
M.~Fagotti,
\newblock Locally quasi-stationary states in noninteracting spin chains,
\newblock SciPost Phys. {\bf 8}(3) (2020).

\bibitem{bastianello2020noise}
A.~Bastianello, J.~De~Nardis, and A.~De~Luca,
\newblock Generalized hydrodynamics with dephasing noise,
\newblock Phys. Rev. {\bf B102}(16) (2020).

\bibitem{CaoTherm18}
X.~Cao, V.~B. Bulchandani, and J.~E. Moore,
\newblock Incomplete Thermalization from Trap-Induced Integrability Breaking:
  Lessons from Classical Hard Rods,
\newblock Phys. Rev. Lett. {\bf 120}(16) (2018).

\bibitem{vas19}
A.~J. Friedman, S.~Gopalakrishnan, and R.~Vasseur,
\newblock Diffusive hydrodynamics from integrability breaking,
\newblock Phys. Rev. {\bf B101}(18) (2020).

\bibitem{DurninTherma2020}
J.~Durnin, M.~J. Bhaseen, and B.~Doyon,
\newblock Non-Equilibrium Dynamics and Weakly Broken Integrability,
\newblock arXiv:2004.11030  (2021).

\bibitem{chippy}
M.~Schemmer, I.~Bouchoule, B.~Doyon, and J.~Dubail,
\newblock Generalized Hydrodynamics on an Atom Chip,
\newblock Phys. Rev. Lett. {\bf 122}, 090601 (2019).

\bibitem{cradle}
J.-S. Caux, B.~Doyon, J.~Dubail, R.~Konik, and T.~Yoshimura,
\newblock Hydrodynamics of the interacting Bose gas in the Quantum Newton
  Cradle setup,
\newblock SciPost Phys. {\bf 6}(6) (2019).

\bibitem{ourunstable}
O.~A. Castro-Alvaredo, C.~De~Fazio, B.~Doyon, and F.~Ravanini,
\newblock On the Hydrodynamics of Unstable Excitations,
\newblock JHEP 2020 {\bf 45} (2020).

\bibitem{tails}
O.~A. Castro-Alvaredo, C.~De~Fazio, B.~Doyon, and A.~A. Zi\'o\l{}kowska,
\newblock Tails of Instability and Decay: a Hydrodynamic Perspective,
\newblock arXiv: 2103.03735  (2021).

\bibitem{roaming}
A.~Zamolodchikov,
\newblock {Resonance factorized scattering and roaming trajectories},
\newblock J.Phys. {\bf A39}, 12847--12862 (2006).

\bibitem{DOREY_1993}
P.~Dorey and F.~Ravanini,
\newblock Staircase models from affine Toda field theory,
\newblock Int. J. Mod. Phys. {\bf 08}(A05), 873--893 (1993).

\bibitem{GenDR} P.~Dorey and F.~Ravanini, Generalizing the staircase models, Nucl.Phys. {\bf B406} 708-726 (1993).

\bibitem{toda2}
A.~Arinshtein, V.~Fateev, and A.~Zamolodchikov,
\newblock Quantum S-matrix of the (1 + 1)-dimensional Toda chain,
\newblock Phys. Lett. {\bf B87}, 389--392 (1979).

\bibitem{toda1}
A.~Mikhailov, M.~Olshanetsky, and A.~Perelomov,
\newblock Two-dimensional generalized Toda lattice,
\newblock Comm. Math. Phys. {\bf 79}, 473--488 (1981).

\bibitem{SSG2}
I.~Arafeva and V.~Korepin,
\newblock Scattering in two-dimensional model with Lagrangian $L=1/\gamma
  (1/2(\partial_\mu u u)^2 + m^2(\cos u-1))$,
\newblock Pis'ma Zh. Eksp. Teor. Fiz. {\bf 20}, 680 (1974).

\bibitem{SSG3}
S.~Vergeles and V.~Gryanik,
\newblock Two-dimensional quantum field theories having exact solutions,
\newblock Yad. Fiz. {\bf 23}, 1324--1334 (1976).

\bibitem{SSG}
B.~Schroer, T.~Truong, and P.~Weisz,
\newblock Towards an explicit construction of the sine-Gordon theory,
\newblock Phys. Lett. {\bf B63}, 422--424 (1976).

\bibitem{za}
A.~Zamolodchikov and A.~Zamolodchikov,
\newblock Factorized S-matrices in two-dimensions as the exact solutions of
  certain relativistic quantum field models,
\newblock Annals Phys. {\bf 120}, 253--291 (1979).

\bibitem{Mossel}
J.~Mossel and J.-S. Caux,
\newblock Generalized TBA and generalized Gibbs,
\newblock J. Phys. {\bf A45}, 255001 (2012).

\bibitem{FM}
D.~Fioretto and G.~Mussardo,
\newblock Quantum quenches in integrable field theories,
\newblock New J. Phys. {\bf 12}(5), 055015 (2010).

\bibitem{DDG} D.X.~Horv\'ath, P.E.~Dorey and G.~Tak\'acs, Roaming form factors for the tricritical to critical Ising flow, JHEP {\bf 07} (2016) 051.

\bibitem{ZAMOLODCHIKOV1991524}
A.~Zamolodchikov,
\newblock From tricritical Ising to critical Ising by thermodynamic Bethe
  ansatz,
\newblock Nucl. Phys. {\bf B358}(3), 524--546 (1991).

\bibitem{ZAMOLODCHIKOV1991497}
A.~Zamolodchikov,
\newblock Thermodynamic Bethe ansatz for RSOS scattering theories,
\newblock Nucl. Phys. {\bf B358}(3), 497--523 (1991).

\bibitem{onemore}
A.~B. Zamolodchikov,
\newblock {TBA equations for integrable perturbed $SU(2)_k \times SU(2)_l /
  SU(2)_{k+ l}$ coset models},
\newblock Nucl. Phys. {\bf B366}, 122--132 (1991).

\bibitem{restrict1}
A.~Leclair,
\newblock Restricted sine-Gordon theory and the minimal conformal series,
\newblock Phys. Lett. {\bf B230}(1), 103--107 (1989).

\bibitem{restrict2}
D.~Bernard and A.~Leclair,
\newblock Residual quantum symmetries of the Restricted sine-Gordon theories,
\newblock Nucl. Phys. {\bf B340}(2), 721--751 (1990).

\bibitem{sovzam}
A.~B. Zamolodchikov,
\newblock {Renormalization Group and Perturbation Theory Near Fixed Points in
  Two-Dimensional Field Theory},
\newblock Sov. J. Nucl. Phys. {\bf 46}, 1090 (1987).

\bibitem{CarLud}
A.~W. Ludwig and J.~L. Cardy,
\newblock Perturbative evaluation of the conformal anomaly at new critical
  points with applications to random systems,
\newblock Nucl. Phys. {\bf B285}, 687--718 (1987).

\bibitem{DXHAk} D\'avid X. Horv\'ath, Hydrodynamics of massless integrable RG flows and a non-equilibrium c-theorem, JHEP {\bf 2019} 10 (2019).

\bibitem{elastic}
T.~Klassen and E.~Melzer,
\newblock {Purely Elastic Scattering Theories and their Ultraviolet Limits},
\newblock Nucl.Phys. {\bf B338}, 485--528 (1990).

\bibitem{kuniba}
A.~Kuniba, T.~Nakanishi, and J.~Suzuki,
\newblock {T-systems and Y-systems in integrable systems},
\newblock J.Phys. {\bf A44}, 103001 (2011).

\bibitem{video} M. Mazzoni, O. Pomponio, O.A. Castro-Alvaredo and F. Ravanini, Staircase Model: Spectral Densities and Effective Velocity, \url{https://youtu.be/rI6KTxLZz_8}.

\end{thebibliography}
\end{document}